\documentclass[a4paper,11pt]{article}
\pdfoutput=1
 \usepackage{jheppub} 
 \usepackage{graphicx}
\usepackage{amsmath}
 \usepackage{amssymb}
 \usepackage{slashed}
\usepackage{dcolumn}
\usepackage{bm}
\usepackage{color}
\usepackage{multirow}

\allowdisplaybreaks

 \newcommand{\lsim}{{\;\raise0.3ex\hbox{$<$\kern-0.75em\raise-1.1ex\hbox{$\sim$}}\;}}
\newcommand{\gsim}{{\;\raise0.3ex\hbox{$>$\kern-0.75em\raise-1.1ex\hbox{$\sim$}}\;}}
\def\bea{\begin{eqnarray}}
\def\eea{\end{eqnarray}}
\def\bec{\begin{center}}
\def\ec{\end{center}}

\def\beq{\begin{equation}}
\def\eeq{\end{equation}}

\def\bea{\begin{eqnarray}}
\def\eea{\end{eqnarray}}
\def\beq#1\eeq{\begin{align}#1\end{align}}
\def\beqnn#1\eeq{\begin{align*}#1\end{align*}}
\def\ba{\begin{array}}
\def\ea{\end{array}}
\def\bc{\begin{center}}
\def\ec{\end{center}}

\newcommand{\dis}[1]{\begin{equation}\begin{split}#1\end{split}\end{equation}}

\preprint{CTPU-PTC-20-28}

\title{Recent  Progress in the Physics of Axions and Axion-Like Particles}
 
 \author{
    Kiwoon Choi\footnote{Electronic address: kchoi@ibs.re.kr}, Sang Hui Im\footnote{Electronic address: imsanghui@ibs.re.kr}, and Chang Sub Shin\footnote{Electronic address: csshin@ibs.re.kr}
    }
     
\affiliation{
 Center for Theoretical Physics of the Universe,  Institute for Basic Science, Daejeon 34126,\\
  South Korea 
    }

\abstract{
The axion is a light pseudoscalar particle postulated to solve issues with the Standard Model, including the strong CP problem and the origin of dark matter. In recent years, there has been remarkable progress in the physics of axions in several directions. An unusual type of axion-like particle termed the relaxion was proposed as a new solution to the weak scale hierarchy problem. There are also new ideas for laboratory, astrophysical, or cosmological searches for axions; such searches can probe a wide range of model parameters that were previously inaccessible. On the formal theory side, the weak gravity conjecture indicates a tension between quantum gravity and a trans-Planckian axion field excursion. Many of these developments involve axions with hierarchical couplings. In this article, we review recent progress in axion physics, with particular attention paid to hierarchies between axion couplings. We emphasize that the parameter regions of hierarchical axion couplings are the most accessible experimentally. Moreover, such regions are often where important theoretical questions in the field are addressed, and they can result from simple model-building mechanisms.
}

\keywords{axions, axion-like particles, axion couplings, axion scales, axion cosmology, axion detection, axion landscape}

\vspace{4cm}

\begin{document} 
\maketitle
\flushbottom
 
\section{Introduction}

Axions and axion-like particles (ALPs) are among the most compelling candidates for
physics beyond  the Standard Model (SM) of particle physics \cite{0807.3125, 1510.07633, 2003.01100}.
They often have good physics motivations and also naturally arise in fundamental theories such as string theory \cite{hep-th/0605206, 0905.4720}.   
In some cases, axion refers to a specific type of pseudo-Nambu-Goldstone  boson  designed to solve the strong CP problem \cite{Peccei:1977hh, Weinberg:1977ma, Wilczek:1977pj}. In this article, we refer to such an axion as a QCD axion, and use the term {\it axions} (and sometimes the term ALPs) for generic pseudo-Nambu-Goldstone bosons associated with non-linearly realized approximate global $U(1)$ symmetries. 

Many different types of axions have been discussed in particle physics and cosmology. 
Some of them are introduced to solve the so-called naturalness problems. The most well-known example is a QCD axion that solves the strong CP problem \cite{Peccei:1977hh, Weinberg:1977ma, Wilczek:1977pj}. Another example is an axion for cosmic inflation \cite{Freese:1990rb},
which would solve the naturalness problems of the Big-Bang cosmology while avoiding unnatural fine tuning in the underlying UV theory.
Recently an unusual type of axion-like particle called the relaxion has been proposed as a new solution to the weak scale hierarchy problem \cite{1504.07551}. Regardless of their role for the naturalness problems, light axions  are compelling candidate for the dark matter in our Universe \cite{Preskill:1982cy, Abbott:1982af, Dine:1982ah,1201.5902}.
Although the data are not convincing enough, a few astrophysical anomalies might be explained by axions with certain specific masses and couplings \cite{1302.1208, 1605.06458, 1704.05189}.

In recent years, there has been significant progress in the physics of axions in several different directions. Such developments include the relaxion solution to the weak scale hierarchy problem \cite{1504.07551}, which has  a variety of interesting phenomenological implications \cite{1610.00680,1610.02025,2004.02899}. 
New ideas for axion searches in laboratory experiments have been proposed \cite{1602.00039, 1801.08127, 2003.02206},
and such searches can probe a wide range of axion masses and couplings that were not accessible before. There are also studies on  the gravitational probe of ultralight axion dark matter \cite{1904.09003}, as well as the axion superradiance from black holes
\cite{0905.4720}.
In addition to these, new theory concepts have generated both constraints and model-building ideas. On one hand, 
it is argued that quantum gravity provides a non-trivial lower bound on axion couplings \cite{hep-th/0601001}, which might be in conflict with the requirement in some models of inflation for trans-Planckian axion field excursions \cite{1412.3457, 1503.00795,1503.03886,1503.04783,1608.06951}.  On the other hand, mechanisms have been presented to naturally produce large hierarchies in axion couplings  \cite{hep-ph/0409138, hep-th/0507205, 1404.6209, 1511.00132,  1511.01827, 1610.07962, 1611.09855, 1709.06085}.

Many of these developments suggest
that the landscape of axion models is much broader than commonly realized. 
Of particular interest are regions where the axion couplings have large hierarchies. As will be discussed in more detail,
those regions are generally the most accessible experimentally,  and often regions where the axion field addresses important theoretical question. Actually,
 large hierarchies among axion couplings are not unexpected. They are technically natural and also can result from simple model-building mechanisms.

This article is organized as follows. In Sec.~\ref{sec:coupling_scale}, we introduce the relevant axion couplings and scale parameters, and present the
observational constraints on those parameters, including the projected sensitivity limits of the planned experiments. In Sec.~\ref{sec:hierarchy}, we present some examples of  the well-motivated axion coupling  hierarchies 
and discuss the model building attempts to generate those hierarchies in low energy effective theory. Sec.~\ref{sec:conclusion} is the summary and conclusion.

\section{Axion couplings and scales \label{sec:coupling_scale}}



Axions are periodic scalar fields, so characterized by a scale
 $f_a$  called the axion decay constant which defines the full field range of the canonically normalized axion:
\bea
\label{decay_const}
a \cong a +2\pi f_a.\eea
Such axions may originate from the phase of complex scalar field \cite{Kim:1979if, Shifman:1979if, Dine:1981rt, Zhitnitsky:1980tq} as 
\bea \label{axion_field}
\sigma = \rho e^{i a/f_{a}} \quad \left(f_a=\sqrt{2}\langle \rho\rangle\right),\eea
or the zero mode of a $p$-form gauge field $A^{(p)}$ in models with extra dimension such as  compactified string theories\footnote{For $p\geq 2$, there can be additional axions  originating from the component
$A^{(p)}_{[\mu\nu m_1..m_{p-2}]}$.} \cite{Witten:1984dg, hep-ph/9902292, hep-th/0303252, hep-th/0605206}:
\bea \label{axion_pform} 
\hskip -0.3cm 
A^{(p)}_{[m_1m_2.. m_p]}(x,y)=
{a(x)} \omega^{(p)}_{[m_1m_2.. m_p]}(y),
\eea
where $x^\mu$ and $y^m$ are the cooridnates of the 4-dimensional Minkowski spacetime and the compact internal space $Y$, respectively, and $\omega^{(p)}$ is a harmonic $p$-form in $Y$.
In the latter case, the axion periodicity  is assured by the Dirac quantization of the axionic string which is a low energy remnant of  the brane which couples (electrically or magnetically) to $A^{(p)}$ in the underlying UV theory.  In that description,
the gauge equivalence of $A^{(p)}$ on a $p$-cycle in $Y$,  which determines the value of $f_a$, is fixed by the couplings and scales involved in the compactification \cite{Choi:1985je, hep-th/0303252, hep-th/0605206}.


Regardless of their UV origin,  axions  can be naturally light if the theory admits an approximate 
 $U(1)$ 
symmetry realized as a shift of the axion field:
\bea
\label{shift_symmetry}
U(1)_{\rm PQ}: \,\,\, {a(x)}\,\,\rightarrow\,\, {a(x)}+  \mbox{constant}.\eea
 This is called the Peccei-Quinn (PQ) symmetry. For $a(x)$ originating from the phase of  complex scalar field, $U(1)_{\rm PQ}$ may arise as an accidental symmetry of the low energy effective theory.  
For $a(x)$ from $p$-form  gauge field, it is the low energy remnant of 
the $p$-form gauge symmetry  $A^{(p)}\rightarrow A^{(p)}+dC^{(p-1)}$ in the underlying higher-dimensional theory.

A key parameter for axion physics is the PQ-breaking coupling $g_{a\Lambda}$ that generates the leading $a$-dependent terms in the axion potential. 
 The corresponding potential can often  be approximated  by a sinusoidal function:
\bea
\label{axion_potential}
V(a)\simeq -\Lambda^4\cos (g_{a\Lambda}{a}).\eea
This can also be written as 
\bea
\label{potential_1}
V(a) \simeq -\Lambda^4\cos \Big( N_{\rm DW}\frac{a}{f_a}\Big),\eea
where $N_{\rm DW}$ is the number of (approximately) degenerate vacua found  over the full field range $2\pi f_a$, which is called the domain wall number. Note that
the coupling $g_{a\Lambda} = N_{\rm DW}/f_a$ also defines  the field range over which the axion potential is
\emph{monotonically} changing:
\bea
\label{mono_range}
\Delta a \equiv  \frac{\pi}{g_{a\Lambda}}= \pi\frac{ f_a}{N_{\rm DW}},\eea
which may set an upper bound on the possible cosmological excursion of the axion field.


\subsection{Axions in cosmology\label{sec:axion_cosmology}}

Axions can play many important roles in cosmology \cite{1510.07633}.
Here we  present three examples which are relevant for our later discussion of axion coupling hierarchies.  


\subsubsection{Axion inflation\label{sec:axion_inf}}
Axion field with a trans-Planckian  $f_a/N_{\rm DW} > M_P\simeq 2.4
\times 10^{18}$ GeV 
can play the role of  inflaton  in the model of natural inflation \cite{Freese:1990rb}.
For the inflaton potential given by Eq.~(\ref{potential_1}),
one finds
\bea
\epsilon \sim \eta \sim \frac{1}{N_e}\sim \left(\frac{M_P}{f_a/N_{\rm DW} }\right)^2,\eea
where $\epsilon=M_P^2(\partial_a V/V)^2/2$ and $\eta =M_P^2(\partial_a^2 V/V)$ are the slow roll parameters, and 
 $N_e$ is the number of  e-foldings.
  Then the observed CMB power spectrum implies $m_a\sim {\cal H}_{\rm inf}/\sqrt{N_e}\sim 10^{13}$ GeV, where ${\cal H}_{\rm inf}\sim \Lambda^2/M_P$  is the inflationary Hubble scale \cite{1807.06211, Freese:1990rb}. This model of inflation is particularly interesting as it predicts the primordial gravitational waves with a strength comparable to the present observational bound \cite{1807.06211, 1510.07633}.

\subsubsection{Axion dark matter\label{sec:axion_dm}}
Light axions are  compelling candidate for dark matter.
The most straightforward way to produce axion dark matter would be the misalignment mechanism \cite{Preskill:1982cy, Abbott:1982af, Dine:1982ah,1201.5902}.
In the early Universe,  the initial value of the axion field is generically misaligned from the present vacuum value, which
can be parametrized as \bea
a(t_i)-a(t_0)=  \Theta_{\rm in} \frac{ f_a}{N_{\rm DW}},\eea
where $t_0$ is the present time and $\Theta_{\rm in}$ is an angle parameter in the range $[0,\pi]$. 
Due to the Hubble friction, the axion field 
has a negligible evolution when
 the Hubble expansion rate ${\cal H}(t)\gg m_a(t)$. In the later time $t=t_{\rm osc}$ when ${\cal H}(t_{\rm osc})\sim m_a(t_{\rm osc})$, it begins to oscillate around the present vacuum value and
  the axion energy density  $\rho_a$ subsequently evolves like a matter energy density. 
Taking  the simple harmonic approximation for the axion potential, 
the resulting axion dark matter abundance turns out to be \cite{1201.5902}
\dis{ \label{dm_abun}
\Omega_{a}(t_0)h^2 \,\simeq\,  0.1 \left(\frac{m_a(t_0)}{\textrm{eV}}\right)^{1/2} \left(\frac{m_a(t_0)}{m_{a}(t_{\rm osc})}
\right)^{1/2}\left( \frac{\Theta_{\rm in}f_a/N_{\rm DW}}{3\times10^{11} \,\textrm{GeV}}\right)^2, 
}
where 
$\Omega_a=\rho_a/\rho_c$
($\rho_c = 3M_P^2{\cal H}^2$) and $h={\cal H}(t_0)/(100 {\rm km}\,{\rm s}^{-1} {\rm Mpc}^{-1})\simeq 0.7$.

For the QCD axion, $m_a(t_0) \approx f_\pi m_\pi /(f_a/N_{\rm DW})$ (see Eq.~(\ref{qcd_axion_mass})) and  $m_a(t_{\rm osc})/m_a(t_0)\approx 6\times 10^{-4}({\rm GeV}/T_{\rm osc})^n$ ($n\simeq 4$) for $T_{\rm osc}\gtrsim 150$ MeV \cite{1606.07494}, where
$m_\pi$ and $f_\pi$ are the pion mass and the pion decay constant, respectively, and $T_{\rm osc}$ is the temperature at $t=t_{\rm osc}$. 
Inserting these into Eq.~(\ref{dm_abun}), one finds the following relic abundance of the QCD axion \cite{Preskill:1982cy, Abbott:1982af, Dine:1982ah}:
\bea
\label{qcd_axion_dm}
\left(\Omega_{a}(t_0)h^2\right)_{\rm QCD} \simeq 0.1\, \Theta_{\rm in}^2 \left(\frac{f_a/N_{\rm DW}}{10^{12}\, {\rm GeV}}\right)^{(n+3)/(n+2)} \quad (n\simeq 4).\eea
Another interesting example is an ultralight ALP dark matter  with $m_a(t_{\rm osc})=m_{a}(t_0)$ \cite{1201.5902, 1610.08297}. Unlike the QCD axion, $m_a$ and $f_a/N_{\rm DW}$ for such an ALP can be regarded as independent parameters, which results in the relic abundance
\bea
\label{alp_dm}
\left(\Omega_{a}(t_0)h^2\right)_{\rm ALP} \simeq 0.1 \,\Theta_{\rm in}^2\left(\frac{f_a/N_{\rm DW}}{10^{17}\,{\rm GeV}}\right)^2 \left(\frac{m_a}{10^{-22}\rm eV}\right)^{1/2}. \eea

\subsubsection{Relaxion\label{sec:relaxion}}
In the relaxion scenario, the Higgs boson mass is relaxed down to the weak scale by the cosmological evolution of an axion-like field called the relaxion,  providing a new solution to the weak scale hierarchy problem \cite{1504.07551}. 
The relaxion potential takes the form
\bea
\label{relax_pot}
V&=&V_0(\theta)+\mu_H^2(\theta)|H|^2 + V_{\rm br}(\theta)\nonumber \\
&=&-\Lambda^4\cos (N_{\rm DW}\theta) + \left(\Lambda_H^2+\Lambda_1^2\cos(N_{\rm DW}\theta)\right) 
|H|^2 -\Lambda_{\rm br}^4 (H)  \cos(N_{\rm br}\theta+\alpha ),
\eea
where $\theta=a/f_a\cong\theta+2\pi$ ($N_{\rm DM}, N_{\rm br} \in \mathbb{Z}$), and $H$ is the Higgs doublet field in the Standard Model whose observed vacuum expectation value is given by $v=\sqrt{2}\langle |H|\rangle=246$ GeV.  $\Lambda_H$ ($\gg v$) is the cutoff scale for the Higgs boson mass, and $\Lambda_{\rm br}(H)$ is a Higgs-dependent scale parameter  which becomes non-zero once $H$ develops a non-zero vacuum value.
Here the terms involving $\cos(N_{\rm DW}\theta)$ are generated at high scales around the cutoff scale, 
so naturally, $
\Lambda\sim \Lambda_1\sim \Lambda_H\gg v$.
On the other hand, the barrier potential $V_{\rm br}=-\Lambda_{\rm br}^4(H) \cos (N_{\rm br} \theta + \alpha)$ is generated at a lower scale around or below $v$, so that $\Lambda_{\rm br}\equiv \Lambda_{\rm br}(v) \lesssim v$.

Initially the effective Higgs mass $\mu_H^2(\theta)=
 \Lambda_H^2+\Lambda_1^2\cos(N_{\rm DW}\theta)$ is supposed to have a large value $\mu_H^2(\theta(t_i))\sim \Lambda_H^2 >  0$. 
 But it is subsequently relaxed down to $\mu_H^2(\theta(t_f))\sim -v^2 < 0$ 
 by the relaxion field excursion
 $a(t_i)-a(t_f)\sim f_a/N_{\rm DW}$  
 driven by the potential $V_0$.
 Since a non-zero $V_{\rm br}$ is developed after the relaxion passes through the critical point $\mu_H^2 (\theta)=0$, 
the relaxion is finally stabilized by the competition between the sliding slope $\partial_\theta V_0$ and the barrier slope $\partial_\theta V_{\rm br}$, which requires 
 \dis{
\frac{N_{\rm br}}{N_{\rm DW}} \sim \frac{\Lambda^4}{\Lambda_{\rm br}^4} \gtrsim\, \frac{\Lambda^4_H}{v^4}.
\label{rel_hie_1}
}
For a successful stabilization, the  scheme also requires a mechanism to dissipate away the relaxion kinetic energy. In the original model \cite{1504.07551}, it  is done by the Hubble friction over a long period of inflationary expansion. 
 Some alternative possibilities will be discussed in Sec. \ref{subsec:relaxion_coupling} in connection with the relaxion coupling hierarchy.






\subsection{Axion couplings to the Standard Model} \label{sec:couplings}

To discuss the axion couplings to the SM, it is convenient to use the angular field  \bea
\theta(x)=\frac{a(x)}{f_a}\cong \theta(x) +2\pi\eea
in the field basis for which {\it only} the axion transforms under the PQ symmetry Eq.~(\ref{shift_symmetry}) and all the SM fields are invariant  \cite{Georgi:1986df}. Here we are interested in axions with $m_a< v \ll f_a$. 
We thus start with an effective lagrangian defined at the weak scale,  which could be derived from a
more fundamental theory defined   at higher energy scale.
Then the PQ-invariant part of the lagrangian is  given by
\bea
\label{pq_invariant}
{\cal L}_0=\frac{1}{2}f_a^2 \partial_\mu\theta\partial^\mu \theta +\sum_{\psi} c_\psi\partial_\mu \theta\bar\psi \bar\sigma^\mu\psi,
\eea
where $\psi=(Q,u^c,d^c, L, e^c)$ denote the chiral quarks and leptons in the SM,
and the derivative coupling to the Higgs doublet current $i(H^\dagger \partial_\mu H-\partial_\mu H^\dagger H)$ is rotated away by an axion-dependent $U(1)_Y$ transformation. 
Generically $c_\psi$ can include flavor-violating components, but we will assume that such components are negligble.

The PQ-breaking part includes a variety of non-derivative axion couplings such as 
 \bea
\label{pq_breaking}
\Delta {\cal L}&=& \frac{\theta(x)}{32\pi^2}\left(c_G G^{\alpha\mu\nu}\widetilde G^\alpha_{\mu\nu}+ c_W W^{i\mu\nu}\widetilde W^i_{\mu\nu}+c_B B^{\mu\nu}\widetilde B_{\mu\nu}\right)\nonumber \\
&&-\,V_0(\theta)-\mu_H^2(\theta)|H|^2 + \cdots ,
 \eea
where $F^X_{\mu\nu}=(G^{\alpha}_{\mu\nu}, W^i_{\mu\nu}, B_{\mu\nu})$ are the $SU(3)_c\times SU(2)_W\times U(1)_Y$ gauge field strengths,  $\widetilde F^X_{\mu\nu} =\frac{1}{2}\epsilon_{\mu\nu\rho\sigma}F^{X\rho\sigma}$ are their duals,  and the ellipsis stands for the PQ-breaking
axion couplings to 
the operators with mass-dimension $\geq 4$ other than $F^X_{\mu\nu}\tilde F^{X\mu\nu}$, which will be ignored in the following discussions.
 We assume that the underlying theory does not generate an axion monodromy \cite{0803.3085}, so allows a field basis for which each term in $\Delta {\cal L}$,
 or $e^{iS_{\rm int}}$ of the corresponding action $S_{\rm int}$,  is invariant under $\theta\rightarrow \theta+2\pi$.
In such field basis, $V_0$ and $\mu_H^2$ are $2\pi$ periodic functions of $\theta$, $c_G$ and $c_W$ are integers, and $c_B$ is a rational number. 
Note that, although the first line of Eq.~(\ref{pq_breaking}) includes $a(x)$, this coupling depends only on the derivative of $a(x)$ in perturbation theory since  $F^X_{\mu\nu}\tilde F^{X\mu\nu}$ is a total divergence.
As a consequence, these terms contribute to the renormalization group (RG) running of the derivative couplings to the SM fermions in Eq.~(\ref{pq_invariant}) \cite{Srednicki:1985xd, hep-ph/9306216}, e.g. 
\bea
\label{rg_psi}
\frac{d c_\psi}{d \ln \mu} =  \sum_{X=G,W,B} -\frac{3}{2}\left(\frac{g_X^2}{8\pi^2}\right)^2\mathbb{C}_{X}(\psi)\Big(  c_X-2\sum_{\psi'}
c_{\psi^\prime}\mathbb{T}_{X}(\psi')  \Big)
\eea
at scales above the weak scale, where $\mathbb{C}_{X}(\psi)$ and $\mathbb{T}_{X}(\psi)$
 are the quadratic Casimir
and Dynkin index of $\psi$.
Some axion models predict $c_\psi=0$ at the UV scale \cite{Kim:1979if, Shifman:1979if}. In such models, the low energy values of $c_\psi$ are determined mainly by their RG evolution including Eq.~(\ref{rg_psi})
\cite{Srednicki:1985xd, hep-ph/9306216}.



To examine their phenomenological consequences,
one may scale down the weak scale lagrangians Eq.~(\ref{pq_invariant}) and Eq.~(\ref{pq_breaking}) to lower energy scales.
This procedure is straightforward
at least at scales above the QCD scale. It
 is yet worth discussing shortly  
how the PQ breaking by $\partial_\theta\mu_H^2$ is transmitted to low energy physics, which is particularly relevant for the low energy phenomenology of the relaxion \cite{1610.00680,1610.02025,2004.02899}.
After the electroweak symmetry breaking, 
a nonzero value of  $\partial_\theta\mu_H^2$ results in a PQ and CP breaking Higgs-axion mixing with the mixing angle 
\bea
\sin \theta_{ah} \simeq {\langle \partial_\theta \ln \mu_H (\theta) \rangle \frac{v}{f_a}}.  \label{mix ang}
\eea
It also gives rise to the axion-dependent masses
of the SM fermions and gauge bosons:
\bea
\sum_\Psi m_\Psi(\theta) \bar\Psi\Psi + M_W^2(\theta) W^{\mu+}W^-_\mu+ \frac{1}{2} M_Z^2(\theta) Z^\mu Z_\mu\eea
with
$\partial_\theta \ln M_\Phi (\theta) =\partial_\theta\ln v(\theta)= \partial_\theta \ln \mu_H(\theta) \quad (M_\Phi=m_\Psi, M_W, M_Z)$,
where the Dirac fermion $\Psi $ denotes the SM quarks and charged leptons, and
$v(\theta)=(-\mu_H^2(\theta)/{\lambda_H})^\frac{1}{2}$ is
 the {axion-dependent Higgs vacuum value} with the Higgs quartic coupling
$\lambda_H$ which is independent of the axion field in our approximation.
Then, integrating out the gauge-charged heavy field $\Phi$  leaves an axion-dependent threshold correction to the low energy gauge couplings as
\dis{
\Delta g = \Delta \beta_g (M_\Phi)  \ln(M_\Phi(\theta)/\mu), \label{threshold_ac}}
where $\Delta \beta_g(M_\Phi)$ is the threshold correction to the beta function at the scale $\mu \sim M_\Phi$.

Scaling the theory down to $\mu\sim 1$  GeV,
the axion effective lagrangian is given by
\bea
\hskip -2cm
{\cal L}&=&\frac{1}{2}f_a^2 \partial_\mu\theta\partial^\mu \theta +\sum_{
{\Psi=q, \,\ell}} c_\Psi \frac{\partial_\mu \theta}{2} \bar\Psi \gamma^\mu\gamma_5 \Psi \
+  \frac{\theta(x)}{32\pi^2}\Big( c_G 
 G^{\alpha\mu\nu}\widetilde G^\alpha_{\mu\nu}  + c_\gamma F^{\mu\nu}\widetilde F_{\mu\nu}\Big)
\nonumber \\
&-&V_0(\theta) + \frac{\mu_H^4(\theta)}{4\lambda_H} -\sum_{
{\Psi=q, \, \ell}} m_\Psi(\theta)\bar\Psi\Psi - \frac{1}{4g_{s}^2(\theta)}G^{\alpha\mu\nu} G^\alpha_{\mu\nu}  -\frac{1}{4 e^2(\theta)}F^{\mu\nu}F_{\mu\nu} +\cdots,
\label{eft_1gev}
 \eea
where $q=(u, d, s)$, $\ell=(e,\mu)$, and 
 $g_s^2(\theta), e^2(\theta)$ include the axion-dependent
threshold corrections Eq.~(\ref{threshold_ac}) from the heavy quarks and tau lepton.
Ignoring the RG evolutions,
$c_\Psi$ and $c_\gamma$ are determined by the weak scale parameters $c_\Psi$, $c_{W,B}$ in Eqs.~(\ref{pq_invariant}),(\ref{pq_breaking}) as
\bea
\hskip -0.3cm 
c_u= c_{Q_1}+ c_{u_1^c},  \quad c_{d,s}=c_{Q_{1,2}}+ c_{d_{1,2}^c},  \quad
c_{e,\mu} = c_{L_{1,2}} + c_{e_{1,2}^c},\quad
c_\gamma = c_W + c_B.\eea
In our case, both $\partial_\theta m_\Psi(\theta)$ and $\partial_\theta g^2(\theta)$ originate from $\partial_\theta\mu_H^2(\theta)$, hence  Eq.~(\ref{eft_1gev}) results in
the following PQ and CP breaking couplings 
at $\mu\sim1$ GeV:
\bea
\label{pq_breaking1} && \qquad 
\theta(x) \Big( \sum_{
{\Psi=q, \, \ell}} \bar{c}_\Psi m_\Psi\bar{\Psi} \Psi  + 
\frac{\bar{c}_G }{32\pi^2} G^{\alpha\mu\nu} G^\alpha_{\mu\nu}   +  \frac{ \bar{c}_\gamma}{32\pi^2} F^{\mu\nu}F_{\mu\nu}\Big) \nonumber \\ 
&& \bar{c}_\Psi = - \langle \partial_\theta \ln \mu_H (\theta) \rangle, \quad 
\bar{c}_G = 2\langle \partial_\theta \ln \mu_H (\theta) \rangle , \quad
\bar{c}_\gamma = -\frac{5}{3} \langle \partial_\theta \ln \mu_H (\theta) \rangle. 
\eea

Many of the observable consequences of axions, in particular those of light axions, are determined  by the couplings to hadrons or to the electron at scales below the QCD scale. Such couplings can be derived in principle from the effective lagrangian Eq.~(\ref{eft_1gev}) defined at $\mu\sim 1$ GeV.
In the following, we present the low energy axion couplings relevant for our subsequent discussion of axion phenomenology \cite{0807.3125,2003.01100}. Specifically, we express the relevant {1PI} couplings at low momenta $p< 1$ GeV
in terms of the Wilsonian model parameters defined at the weak scale while ignoring the subleading corrections.
Many of the couplings to hadrons can be obtained by an appropriate matching between Eq.~(\ref{eft_1gev}) and the chiral lagrangian of the nucleons and light mesons \cite{Kaplan:1985dv, Srednicki:1985xd, hep-ph/9306216}. For the PQ and CP breaking axion couplings below the QCD scale, there are two sources in our case. One is the coupling $c_G$ to the QCD anomaly combined with nonzero value of the QCD vacuum angle $\theta_{\rm QCD}=c_G\langle \theta(x)\rangle$, and the other is the  Higgs-axion mixing induced by $\partial_\theta \mu_H^2(\theta)\neq 0$. For the couplings from the Higgs-axion mixing, one can use the known results
for the low energy couplings of the light Higgs boson \cite{Leutwyler:1989tn, Chivukula:1989ze}. 
One then finds
\bea
{{\cal L}_{\rm 1PI}}&=& \frac{g_{a\gamma}}{4} a F^{\mu\nu}\widetilde F_{\mu\nu} +  \frac{\bar g_{a\gamma}}{4} a F^{\mu\nu}F_{\mu\nu}+
\sum_{\Psi=p,n,e,\mu}\left( g_{a\Psi} \frac{\partial_\mu a}{2m_\Psi}\bar \Psi\gamma^\mu\gamma_5 \Psi
 + \bar g_{a\Psi} {a} \bar \Psi \Psi \right)
\nonumber \\
&+& \frac{1}{2} \bar{g}_{a\pi}^{(1)} a \left( \partial_\mu \pi^0 \partial^\mu \pi^0 +2 \partial_\mu \pi^+ \partial^\mu \pi^- \right) +\frac{1}{2} \bar{g}_{a\pi}^{(2)}  a\left( \pi^0 \pi^0 + 2\pi^+ \pi^- \right), \label{1pi_coupling}
\eea
where
\bea
\label{1pi_coupling_list}
g_{a \gamma} &=&  \frac{\alpha_{\rm em}}{2\pi}\frac{1}{f_a} \Big(c_\gamma -\frac{2}{3} \frac{(4m_d+m_u)}{(m_u+m_d)}c_G - 2 \sum_{\ell=e, \mu} \left(1-A_f(\tau_\ell)\right) c_\ell\Big), \nonumber \\
\bar{g}_{a \gamma} &=& \frac{\alpha_{\rm em}}{2\pi}\frac{1}{f_a} \Big(-\frac{11}{9} +\sum_{\Phi=K^\pm, \pi^\pm} \frac{11 + 8 \tau_{\Phi}}{54} \bar A_s(\tau_\Phi) + \frac{4}{3}\sum_{\ell=e, \mu} \bar A_f (\tau_\ell) \Big)  \langle\partial_\theta \ln \mu_H (\theta)\rangle, \nonumber \\ 
g_{ap} &=& \frac{m_p}{f_a}\left(-0.47c_G + 0.88 c_u -0.39 c_d\right), \quad
g_{an} = \frac{m_n}{f_a}\left(-0.02c_G -0.39 c_u +0.88 c_d\right), \nonumber 
\\ 
\bar{g}_{ap} &=& \bar{g}_{an}= \frac{m_N}{f_a}\left(-\frac{2}{9} \langle\partial_\theta \ln \mu_H (\theta)\rangle  + {\theta}_{\rm QCD}\frac{m_u m_d}{(m_u + m_d)^2} \frac{\sigma_{\pi N}}{m_N}\right) , \nonumber \\
g_{a\ell}&=& c_\ell \frac{m_\ell}{f_a}, \quad \bar{g}_{a\ell} = - \langle\partial_\theta \ln \mu_H (\theta)\rangle\frac{m_\ell}{f_a}  \quad (\ell=e,\mu),  \nonumber \\
\bar{g}_{a\pi}^{(1)} &=& \frac{4}{9} \frac{\langle\partial_\theta \ln \mu_H (\theta)\rangle}{f_a}, \quad
\bar{g}_{a\pi}^{(2)} = \frac{m_\pi^2}{f_a}\left(-\frac{5}{3} \langle\partial_\theta \ln \mu_H (\theta)\rangle  + {\theta}_{\rm QCD}  \frac{m_u m_d}{(m_u + m_d)^2}\right).
\eea
Here we use
the  result of \cite{1511.02867} for $g_{aN}$ ($N=p,n$), $\sigma_{\pi N} = \frac{1}{2} (m_u + m_d) \langle N | (\bar{u} u + \bar{d} d) | N \rangle \simeq 42 \, {\rm MeV}$ for  $\bar g_{aN}$ \cite{2007.03319}, and ignore the contribution to $\bar g_{a\gamma}$ from $\theta_{\rm QCD}$.
The $\theta_{\rm QCD}$ contribution to $\bar{g}_{aN}$ is obtained in the isospin symmetric limit \cite{Moody:1984ba} (see \cite{2006.12508} for the subleading corrections).
The 1PI axion-photon
couplings $g_{a\gamma}$ and $\bar g_{a\gamma}$  describe the processes with on-shell photons. 
These couplings depend on the variables  $\tau_{j} = p_a^2/4m_{j}^2$ where $p_a$ ($<1$ GeV) is the axion 4-momentum  and  $m_j$ is the appropriate hadron or lepton mass. For $p_a^2\ll m_j^2$, it is sufficient to approximate the loop functions as $A_f\simeq \bar A_s\simeq \bar A_f\simeq1$; for the loop functions when
$p_a^2\gtrsim m_j^2$, see  \cite{Gunion:1988mf, Leutwyler:1989tn}. 

Let us finally consider the axion effective potential. 
For the weak scale lagrangian  Eq.~(\ref{pq_breaking}),
it is obtained as
\bea
\hskip -0.7cm
\label{V_a_eff}
V_{\rm eff}= V_0(\theta) -\frac{1}{4\lambda_H}\mu_H^4(\theta) - 
\frac{f^2_\pi m_\pi^2}{m_u+m_d}\sqrt{m_u^2+m_d^2+2m_um_d\cos(c_G\theta)}+\cdots.
\eea
The second term of RHS arises from the electroweak symmetry breaking and 
the third term is generated by low energy QCD dynamics through the axion coupling  to the gluon anomaly \cite{1511.02867}. Here the ellipsis denotes the subleading contributions which include for instance
those from the PQ-breaking couplings $\bar c_G$ and $\bar c_q$  in
the effective lagrangian
 Eq.~(\ref{pq_breaking1}).
These are  negligible compared to the electroweak symmetry breaking contribution in our case. 
For the QCD axion, the effective potential is dominated by the term induced by the gluon anomaly. More  
specifically,
$|\partial_\theta V_0|$, $|\partial_\theta \mu_H^4|$ and the ellipsis part are all assumed to be smaller than  $10^{-10} f_\pi^2 m_\pi^2$, so that
the strong CP problem is solved with 
$|\theta_{\rm QCD}|=|c_G{\langle a\rangle}/{f_a}|\lesssim 10^{-10}$.
The corresponding QCD axion mass is found by expanding Eq.~(\ref{V_a_eff}) about $\theta=0$ and is given by
\bea 
\label{qcd_axion_mass}
m_{a_{\rm QCD}}\simeq  c_G\frac{f_\pi m_\pi}{f_a} \frac{\sqrt{m_u m_d}}{m_u+m_d} \simeq 5.7\, \mu{\rm eV} \left(\frac{10^{12} \,{\rm GeV}}{f_a/c_G} \right). \label{QCD axion mass}
\eea



\subsection{Theory  constraints on axion couplings\label{sec:wgc}}

Studies of the quantum properties of black holes and works with string theories suggest that there are non-trivial constraints on effective field theories having a UV completion incorporating  quantum gravity \cite{1903.06239}. For instance, it has been argued that 
 for axions in theories compatible with quantum gravity, there  exist certain instantons which couple to axions with a strength stronger than gravity. This has been proposed as a generalization of the weak gravity conjecture (WGC) on $U(1)$ gauge boson, which states that there exists a particle with mass $m$ and charge $Q$ satisfying $Q\geq m/M_P$ \cite{hep-th/0601001,1402.2287}. 
 Specifically
the axion WGC \cite{hep-th/0601001,1412.3457,1503.00795, 1503.03886,1503.04783}  suggests that
given the canonically normalized  axions $\vec a=(a_1,a_2,\cdots,a_N)$,
there is a set of instantons $\{I\}$
with the Euclidean actions $\{S_I\}$ and  the axion-instanton couplings
\bea
\vec {g}_{aI}= \left(\frac{1}{f_{I1}}, 
\frac{1}{f_{I2}},\dots, \frac{1}{f_{IN}}\right), \eea
which would generate PQ-breaking 
amplitudes (for a constant background axion field) \bea
{\cal A}_I \propto
\exp ({-S_I+i \vec g_{aI}\cdot \vec a}),\eea 
for which the convex hull spanned by 
$M_P\vec {g}_{aI}/S_I$ includes the $N$-dimensional unit ball. 
This convex hull condition can also be expressed as follows.  For an arbitrary linear combination of $N$ axions, e.g. $a= \hat u\cdot \vec a$  with $|\hat u|=1$, there exists an instanton $I$, which we call the WGC instanton, with the axion-instanton coupling  satisfying\footnote{A model-dependent coefficient of order unity can be multiplied to this bound.}
\bea
\label{wgc_bound2}
g_{a{\rm WGC}}\equiv |\hat u\cdot \vec {g}_{aI}|\,\gtrsim \,\frac{S_I}{M_P}.
\eea
For axions from $p$-form gauge fields in string theory, this bound is often saturated by the couplings to the corresponding brane instantons
\cite{Dine:1986zy, hep-ph/9902292, hep-th/0303252, hep-th/0605206}.

To examine the implications of the axion WGC, one sometimes assumes that
the axion-instanton couplings that span  the convex hull satisfying the condition Eq.~(\ref{wgc_bound2}) also generate the \emph{leading} terms in the axion potential \cite{1503.00795,1503.03886, 1503.04783, 1504.00659, 1506.03447, 1607.06814}. 
However, this assumption appears to be too strong to be applicable for generic case.
Generically the axion potential depends on many features of the model other than  the couplings and Euclidean actions of the WGC instantons that are constrained as Eq.~(\ref{wgc_bound2}). 
In this regard, it is plausible
that  some of the leading terms in the axion potential 
are generated by certain dynamics \emph{other than} the WGC instantons,  
e.g. a confining YM dynamics, additional instantons whose couplings are \emph{not} involved in spanning the convex hull for the WGC, or Planck-scale suppressed higher-dimensional operators for accidental $U(1)_{\rm PQ}$ 
 \cite{1412.3457, 1503.04783, 1504.00659, 1608.06951}. 
We therefore adopt the viewpoint that the axion WGC implies the existence of 
certain instantons (called the WGC instantons) whose couplings span the convex hull satisfying the bound Eq.~(\ref{wgc_bound2}) while leaving the dynamical origin of the axion potential as an independent feature of the model.

The axion WGC bound Eq.~(\ref{wgc_bound2}) can be written in a form useful for ultralight axions. For this,
let us parametrize the axion potential induced by the WGC instanton as
\bea
 \label{wgc_potential}
\delta V_{\rm WGC} = M_P^2 \Lambda_{I}^2 e^{-S_{I}}\cos\Big(N_{\rm WGC}\frac{a}{f_a}\Big),\eea
where $\Lambda_{I}$ is a model-dependent scale parameter and $N_{\rm WGC}$ is an integer that characterizes the axion coupling to the WGC instanton, $g_{a{\rm WGC}}= N_{\rm WGC}/f_a$. 
This may or may not be the leading term in the axion potential. At any rate, it provides a lower bound on the axion mass:
\bea
\label{mass_bound}
m_a^2 \,\gtrsim\,  e^{-S_{I}} N_{\rm WGC}^2M_P^2\Lambda_{I}^2/{f^2_a},\eea
implying $S_I\gtrsim 2\ln (\Lambda_I/m_a)$ and therefore
\bea
\label{wgc_bound}
g_{a{\rm WGC}}\,\gtrsim \,\frac{2\ln(\Lambda_I/m_a)}{M_P}.
\eea
One may further assume that
the WGC instanton gives a non-perturbative superpotential in the context of supersymmetry, which happens often in explicit string models 
\cite{Dine:1986zy, 0902.3251, 1610.08297}. This additional assumption leads to 
\bea
\label{wgc_gravitino}
\Lambda_I^2\sim m_{3/2}M_P,\eea
 where $m_{3/2}$ is the gravitino mass.

\subsection{Observational constraints on axion couplings}

Low energy axion couplings are subject to constraints from various laboratory experiments and  astrophysical or cosmological observations. They can be tested also by a number of planned experiments.  In this section
we summarize those constraints and the sensitivities of the planned experiments with the focus on those relevant for axion coupling hierarchies.
More comprehensive review of the related subjects can be found  in \cite{1602.00039, 1801.08127, 1904.09003, 2003.02206}.

\subsubsection{Non-gravitational probes\label{sec:ng_probe}} 

\begin{figure}[th]
\begin{center}
 \begin{tabular}{l}
\includegraphics[width=0.7\textwidth]{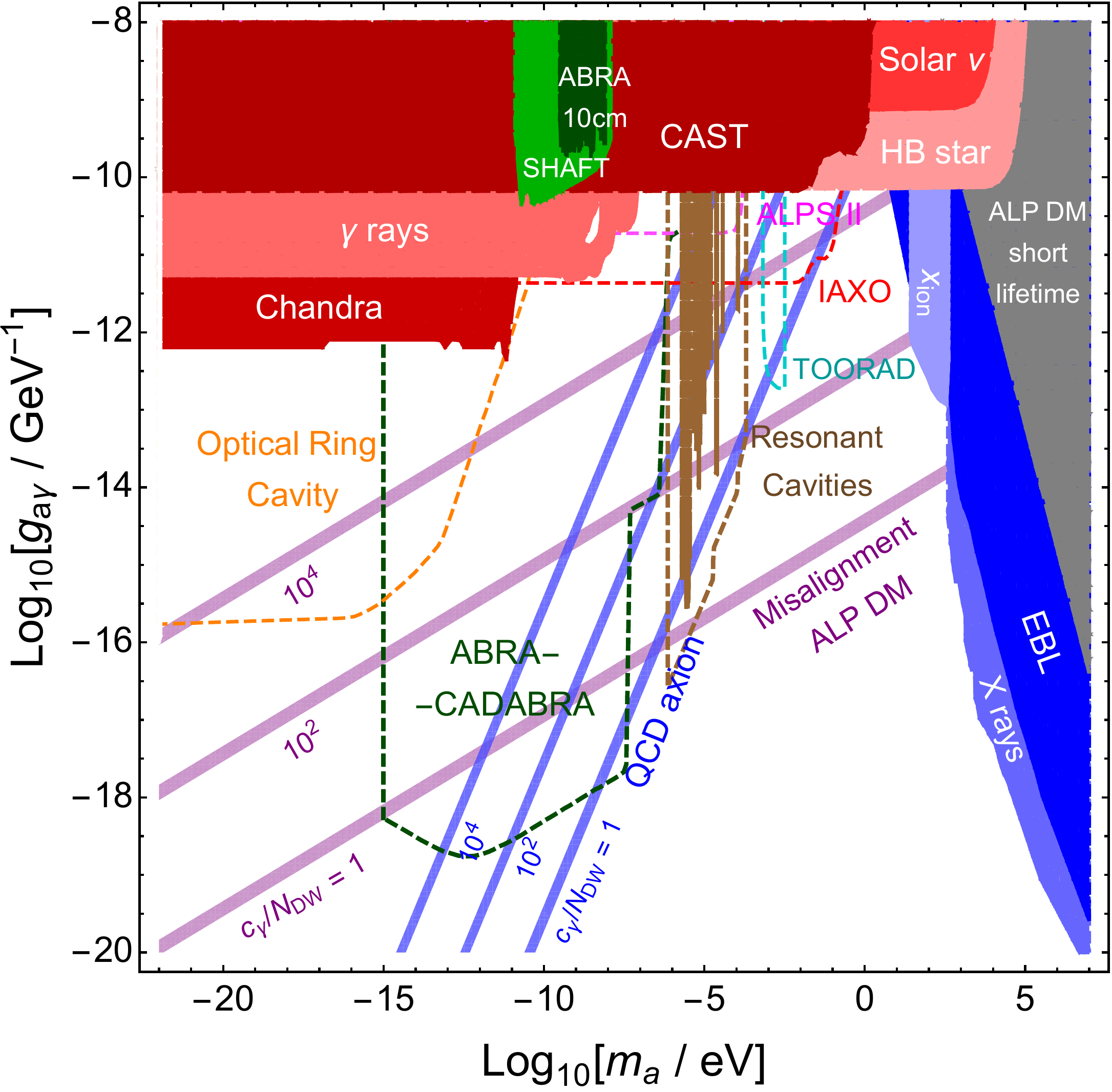} 
   \end{tabular}
  \end{center}
  \caption{Constraints and future probes on the axion-photon coupling $g_{a \gamma}$. The shaded regions are excluded by the existing laboratory, astrophysics and cosmology bounds, and the dashed lines show the sensitivities of the planned experiments. We also depict $g_{a\gamma}$ of  ALP dark matter (pink) and the QCD axion (blue) for three different values of $c_\gamma/N_{\rm DW}$, where $N_{\rm DW}=c_G$ for the QCD axion.}
\label{cp-conv-a}
\end{figure}

\begin{figure}[th]
\begin{center}
 \begin{tabular}{l}
 \includegraphics[width=0.7 \textwidth]{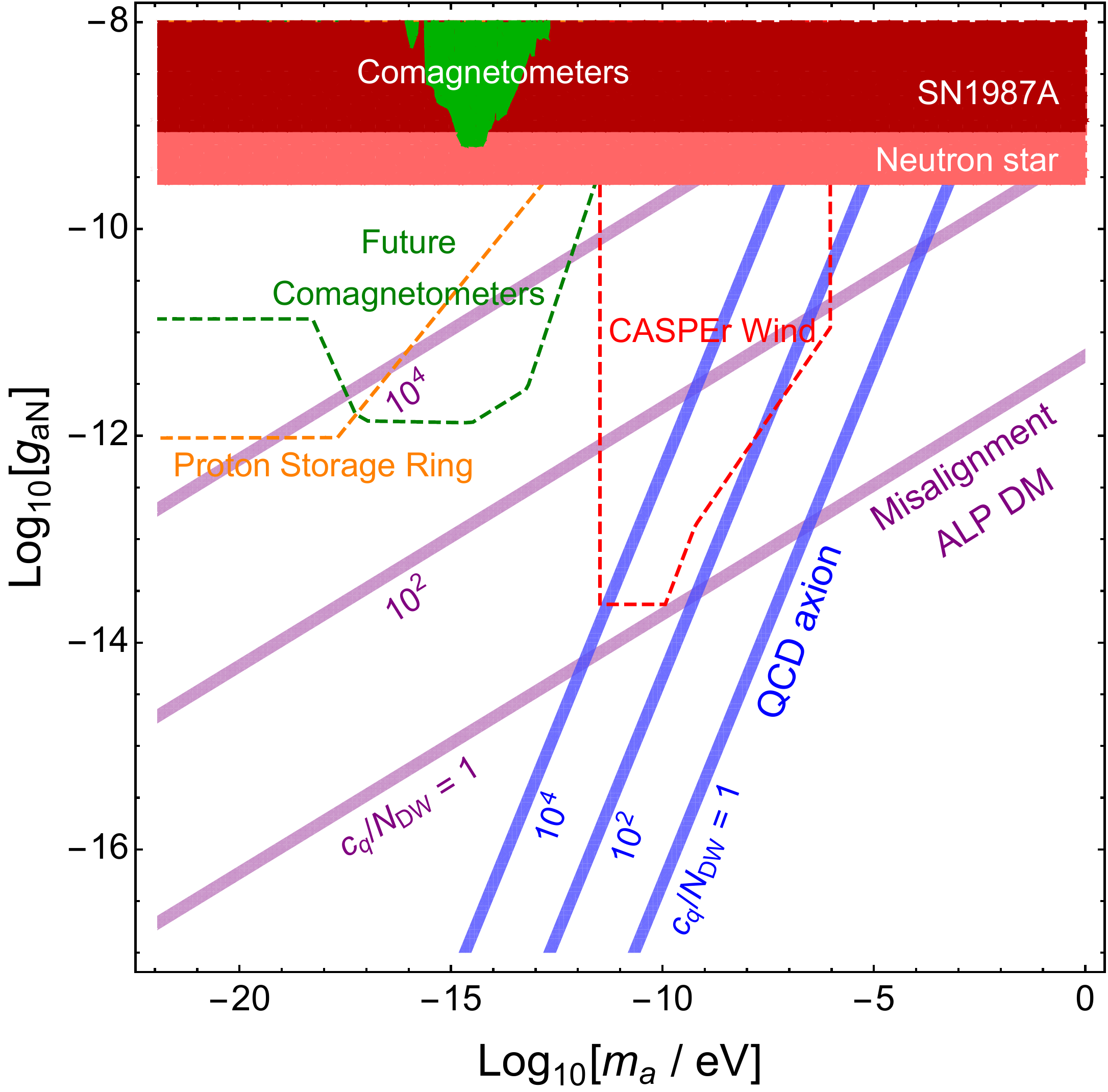}
   \end{tabular}
  \end{center}
  \caption{Constraints and future probes on the axion-nucleon coupling $g_{a N}$. The shaded regions are excluded by the existing laboratory and astrophysics bounds, and the dashed lines show the sensitivities of the planned experiments. We also depict $g_{aN}$ of  ALP dark matter (pink) and the QCD axion (blue) for three different values of $c_q/N_{\rm DW}$, where $N_{\rm DW}=c_G$ for the QCD axion.}
\label{cp-conv-b}
\end{figure}

The CP-conserving axion-photon coupling $g_{a \gamma}$ has been most widely studied for the experimental search for axions.
Fig.~\ref{cp-conv-a}  summarizes the current bounds and future experimental reach for
$g_{a \gamma}$ over the vast range of the axion mass $m_a$. 
 The axion haloscopes (resonant cavities \cite{Sikivie:1983ip, 2010.00169, 1910.11591, 2003.10894}, ABRACADABRA \cite{1602.01086}, Optical Ring Cavity \cite{1805.11753}, TOORAD \cite{1807.08810}, see also \cite{2007.15656} for a recent proposal) are based on the hypothesis
that axions constitute (certain fraction of) the dark matter (DM) in our Universe.
For the axion DM density $\rho_a$, their goal is to detect EM waves arising from the axion-induced effective current
\dis{ 
g_{a \gamma} \partial_t a(t) \vec{B} \approx  g_{a \gamma} \sqrt{2\rho_a} \sin (m_a t) \vec{B}.}
The projected sensitivity limits (dashed lines) in the figure are obtained when  $\rho_a = \rho_{\rm DM} \simeq 0.4 \, \textrm{GeV} /\textrm{cm}^3$.
Otherwise the limits are to be scaled by the factor $\sqrt{\rho_a/\rho_{\rm DM}}$. 
For comparison, we also display  the predicted values of $g_{a\gamma}\sim ({\alpha_{\rm em}}/{2\pi})({c_\gamma}/{f_a})$  for two specific type of  axions with $c_{\gamma}/N_{\rm DW}=1,\,10^2,\,10^4$:
i) ALP DM with $\rho_a=\rho_{\rm DM}$ given by Eq.~(\ref{alp_dm}) with  $\Theta_{\rm in}=1$, which is
 produced by the misalignment mechanism
 (pink lines) and ii) the QCD axion (blue lines).
For the QCD axion, we do not specify its cosmological relic abundance since the corresponding blue lines can be determined by Eq.~(\ref{qcd_axion_mass}) without additional information.
Our result shows that for both ALP DM and the QCD axion, the parameter region more easily accessible by the on-going or planned experiments has 
$c_\gamma/N_{\rm DW}\gg 1$, which parametrizes
the hierarchy between $g_{a\gamma}\sim ({\alpha_{\rm em}}/{2\pi})({c_\gamma}/{f_a})$ 
and the coupling $g_{a\Lambda}=N_{\rm DW}/f_a$  to generate the leading axion potential ($N_{\rm DW}=c_G$ for the QCD axion).

A similar plot for the axion-nucleon  coupling $g_{aN}$ is given in Fig.~\ref{cp-conv-b} ,
including experimental sensitivities and
the predicted coupling for ALP DM and QCD axion with
$c_q/N_{\rm DW}=1,\, 10^2, \, 10^4$. The relevant experiments are CASPEr Wind \cite{1711.08999}, comagnetometers \cite{1709.07852, 1907.03767}, and proton storage ring \cite{2005.11867}.
They are trying to find the axion DM-induced effective magnetic field interacting with the nucleon spin, whose strength is proportional to
\dis{
g_{aN} \vec \nabla a(t, \vec x) \approx g_{aN} \sqrt{2\rho_a} \vec v \sin \left(m_a t - m_a \vec v \cdot \vec x \right), 
}
where $\vec v$ ($|\vec v|\sim 10^{-3}$) is the axion DM virial velocity with respect to the earth. Again its sensitivity limit in the figure is obtained for $\rho_a = \rho_{\rm DM}$, so has to be 
scaled by the factor $\sqrt{\rho_a/ \rho_{\rm DM}}$ otherwise. We see that the ALP parameter region  more easily probed by those experiments has $c_q/N_{\rm DW}\gg 1$, representing the hierarchy between $g_{aN}\sim c_qm_N/f_a$ and
$g_{a\Lambda}=N_{\rm DW}/f_a$. 
Although not shown in the figure, 
recently the QUAX experiment 
has excluded the axion-electron coupling $g_{ae} \gtrsim 2 \times 10^{-11}$ for $m_a \simeq 43\, \mu$eV \cite{2001.08940}, which would correspond to $c_e/N_{\rm DW}\gtrsim 10^{4}$ for the QCD axion.

\begin{figure}[th]
\begin{center}
 \begin{tabular}{l}
  \includegraphics[width=0.7 \textwidth]{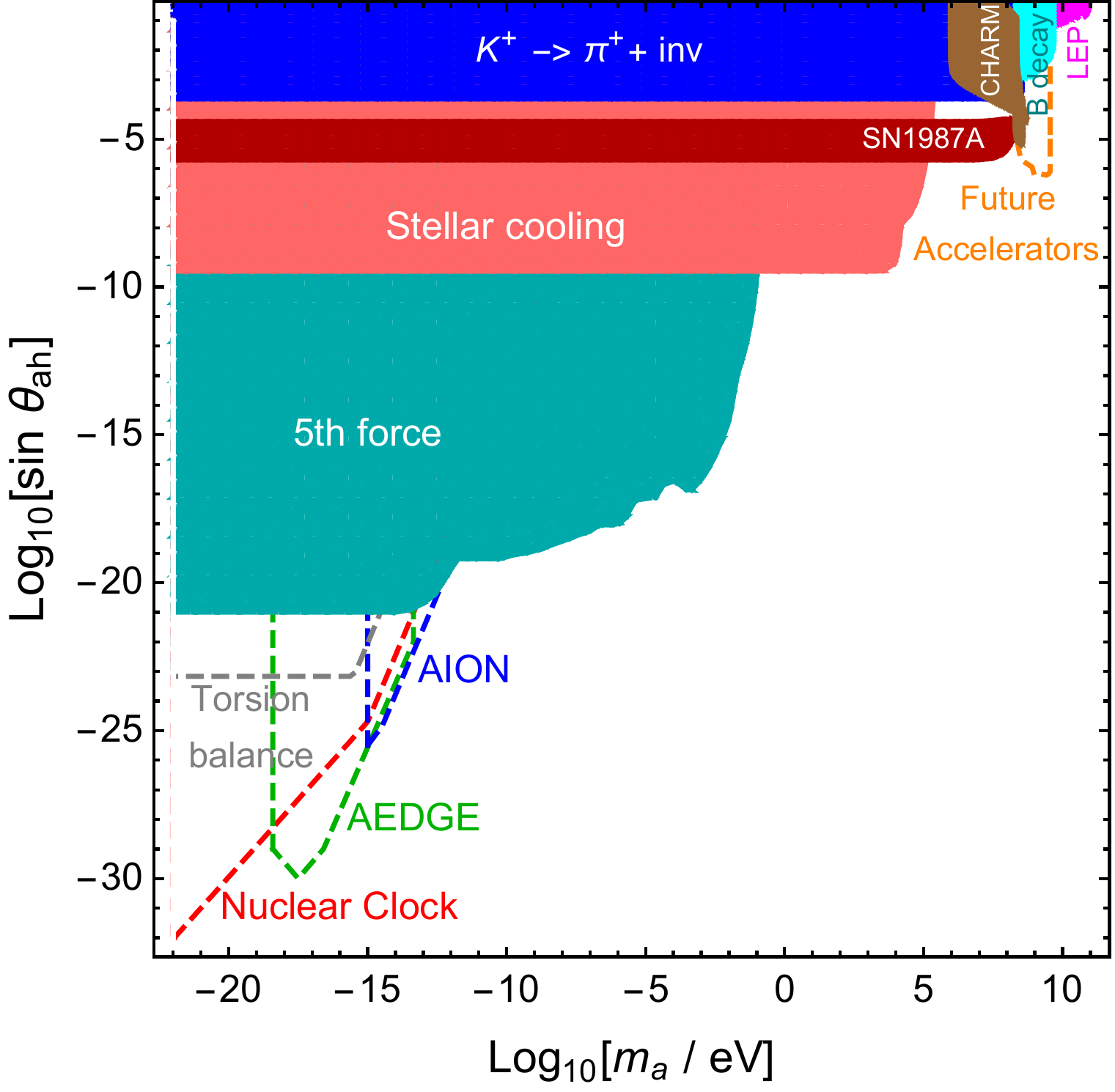}
   \end{tabular}
  \end{center}
  \caption{Constraints on the Higgs-axion mixing $\theta_{ah}$ from those on CP-violating axion couplings. 
  The shaded regions are excluded by the existing constraints, while the dashed lines show the sensitivities of future experiments. 
  }
\label{cpv}
\end{figure}

There are also a number of bounds and experimental probes on 
CP-violating axion couplings.
As described in Sec. \ref{sec:couplings}, 
such couplings can be induced dominantly by the Higgs-axion mixing Eq.~(\ref{mix ang}), which is indeed the case for the relaxion \cite{1610.00680,1610.02025,2004.02899}.
We thus summarize in  Fig.~\ref{cpv}  the available constraints and future prospect on CP-violating axion couplings in terms of the Higgs-axion mixing angle $\theta_{ah}$. 
Exhaustive reference lists for them can be found in \cite{2004.02899, 1610.00680, 1610.02025, 2010.03889}.
For axions lighter than MeV scale,
the constraints are from the axion-mediated 5th force induced by $\bar{g}_{aN}$,
stellar cooling by $\bar{g}_{ae}$ \cite{1611.05852}, and supernova (SN1987A) cooling
by $\bar{g}_{aN}$. The currently  unconstrained supernova trapping window between 100 keV and 30 MeV may be explored by the GANDHI experiment \cite{1810.06467}. {Ultralight axion dark matter can be tested by a future nuclear clock experiment \cite{2004.02899}, torsion balances \cite{1512.06165}, and atom interferometers such as AION \cite{1911.11755} and AEDGE \cite{1908.00802} through its CP-violating couplings}. These experiments will probe axion DM-induced oscillations of fundamental constants like the electron mass, the nucleon mass, and the fine structure constant via the CP-violating couplings in  Eq.~(\ref{1pi_coupling}). So their sensitivities are proportional to the background axion DM field $a(t,x) \approx \sqrt{2\rho_a}/m_a \cos(m_a t)$. The sensitivity lines in the figure are again obtained for $\rho_a = \rho_{\rm DM}$. 
On the other hand, axions heavier than MeV scale are constrained by CHARM beam dump experiment, rare meson decays ($K\rightarrow \pi + a \,[a\rightarrow \textrm{inv}]$, $B\rightarrow K + a \,[ a\rightarrow \mu \mu]$), and LEP ($e^+ e^- \rightarrow Z \rightarrow Z a$). Heavy axions around GeV scale are to be probed by various future accelerator experiments searching for long-lived particles such as FASER, CODEX-b, SHiP, and MATHUSLA \cite{1710.09387}.
For the QCD axion, the dominant source of CP violation is a non-zero $|\theta_{\rm QCD}| < 10^{-10}$, which might be too tiny to be probed by the current experiments. 
Yet ARIADNE experiment \cite{1403.1290, 1710.05413} plans to probe
$\theta_{\rm QCD}$ several orders of magnitude below $10^{-10}$ by observing the axion-mediated 
monopole-dipole force $\propto \bar{g}_{aN} g_{aN}$.


  \subsubsection{Gravitational probes} 
 
\begin{figure}[th]
\begin{center}
 \begin{tabular}{l}
  \includegraphics[width=0.7 \textwidth]{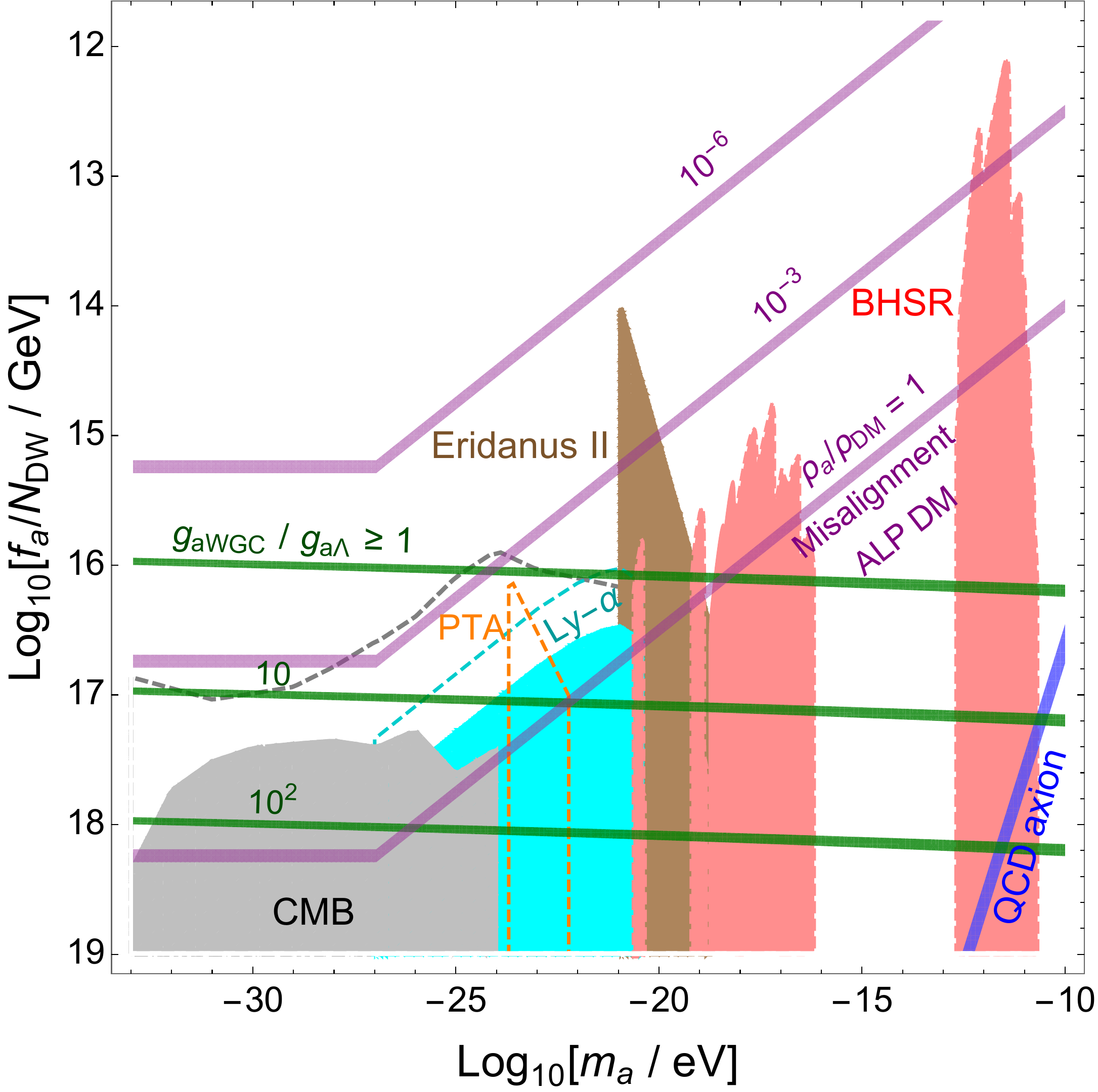}
   \end{tabular}
  \end{center}
  \caption{Constraints on the axion scale $f_a/N_{\rm DW}$ from gravitational probes.
  The shaded regions are excluded by the existing constraints, while the dashed lines show the sensitivities of future experiments. 
 Here $f_a/N_{\rm DW}$ is identified as the
 the field excursion $\Delta a$ for ALP DM or DE, and the axion self-interaction scale for BHSR.  
  We also show the axion coupling hierarchy implied by the weak gravity conjecture (green).}
\label{gravity}
\end{figure}

Gravitational constraints provide a complementary probe of axions. Whereas the non-gravitational probes described in Sec.~\ref{sec:ng_probe} constrain axions with relatively heavier masses and stronger couplings, gravitational probes can constrain extremely light axions with large,  
nearly Planckian, values of $f_a/N_{\rm DW}$.
Such ultralight axions  may constitute a substantial amount of dark energy (DE) or dark matter (DM)
as suggested by Eq.~(\ref{alp_dm}) while having a large  de Broglie wavelength
which can have significant cosmological and astrophysical implications. The relevant axion mass range can be classified by three windows: i) DE-like window $m_a \lesssim 10^{-27}$ eV, ii) ultralight DM  window $10^{-27}\, \textrm{eV} \lesssim m_a \lesssim 10^{-19} \, \textrm{eV}$, and iii) black hole (BH) superradiance  window $10^{-21}\, \textrm{eV} \lesssim m_a \lesssim 10^{-11} \, \textrm{eV}$, which has some overlap with ii).
In Fig.~\ref{gravity}, we depict the existing constraints and expected future limits from 
gravitational probes of ultralight DE-like or DM
axions produced by the initial misalignment $\Delta a =f_a/N_{\rm DW}$, as well as the regions excluded by the black hole (BH) axion superradiance.

For the DE-like window $m_a \lesssim 10^{-27}$ eV, the axion field begins to oscillate after the matter-radiation equality and acts as an early DE component. 
As a result, the locations of the cosmic microwave background (CMB) acoustic peaks shift to larger angular scales (lower $\ell$) and the Universe gets younger. It also increases the largest scale anisotropies through the Integrated Sachs-Wolfe (ISW) effect \cite{1410.2896}.  These constrain the amount of axion by the CMB observations as depicted in Fig.~\ref{gravity}.  For $m_a \lesssim 10^{-32}$ eV, the effect of axion on CMB becomes almost indistinguishable from the cosmological constant within the current precision, so the constraint is weakened.  If $m_a \lesssim 10^{-33}$ eV, the axion field rolls slowly to this day and behaves like the standard DE.

For the ultralight DM  window $10^{-27}\, \textrm{eV} \lesssim m_a \lesssim 10^{-19} \, \textrm{eV}$, 
the axion oscillates before the matter-radiation equality, but  has a cosmic size of 
de Broglie wavelength which would affect cosmic structure formation. There are a number of constraints on this mass range from CMB \cite{1708.05681, 1806.10608, 2003.09655}, pulsar timing array (PTA) \cite{1810.03227}, Lyman-$\alpha$ forest \cite{1708.00015, 1611.00036}, and ultra-faint dwarf galaxy (Eridanus II) \cite{1810.08543}, which are depicted in Fig.~\ref{gravity}.
If $10^{-27}\, \textrm{eV} \lesssim m_a \lesssim 10^{-25} \, \textrm{eV}$, the height of the CMB acoustic peaks becomes higher than that of the $\Lambda$CDM because the axion behaves like DE until the time close to the recombination. Moreover, the wave-like property of axion DM suppresses the growth of density perturbation below a certain comoving Jeans scale \cite{astro-ph/0003365, 1510.07633}, 
affecting the gravitational lensing of CMB \cite{astro-ph/9810092}.
Those effects are undiscerned for $m_a \gtrsim 10^{-25}$ eV with the current data precision, lifting the CMB constraint.  The effects of axion DM on small scales are still significantly constrained by Lyman-$\alpha$ forest data \cite{1708.00015, 1611.00036} and by the evolution of dwarf galaxy Eridanus II \cite{1810.08543}, extending the excluded region up to $m_a \lesssim 10^{-19}$ eV. While PTA currently puts only a weak constraint on the time-oscillating pressure of axion DM, it may eventually probe the dashed region with a 10 year data set \cite{1810.03227}. 

For the BH superradiance window $10^{-21}\, \textrm{eV} \lesssim m_a \lesssim 10^{-11} \, \textrm{eV}$, 
observations of spinning black holes and gravitational waves (GWs) can constrain the existence of axions \cite{0905.4720, 1004.3558}. Let us shortly discuss  some of the underlying physics for this issue (for the details, see \cite{1501.06570}). \emph{Superradiance} is a phenomenon in which incident waves (or bosonic particles) are amplified by extracting energy and angular momentum from a rotating medium.
For axion superradiance, a rotating BH provides such a medium. Moreover, because of the gravitational attraction, the emitted axions can form a bound state around the BH, called the axion cloud. This provides a continuous source of incident axions, and can cause an exponential growth of the axion cloud  by extracting a substantial fraction of the angular momentum of the BH.
In fact, such an amplification of the axion cloud is efficient only when the size of the axion cloud $r_a=(\alpha_g m_a)^{-1}$ ($\alpha_g \equiv G_NM_{\rm BH} m_a$)  is comparable to the BH size $r_{\rm BH}=G_NM_{\rm BH}$, so only for
the axion mass window $m_a \sim (0.1-1)r_{\rm BH}^{-1}$.  Therefore, the observations of highly spinning stellar black holes with $M_{\rm BH}={\cal O}(10 M_\odot)$,  the supermassive black holes with $M_{\rm BH} = {\cal O}(10^6{\rm -}10^8 M_\odot)$, and the recently observed spinning M87$^*$ with the mass $6.5\times 10^9\,M_\odot$ provide the strong constraints on the existence of axions in the mass ranges of  $10^{-13}{\rm-}10^{-11}\,{\rm eV}$, $10^{-19}{\rm-}10^{-16}\,{\rm eV}$, and $10^{-21}{\rm-}10^{-20}\,{\rm eV}$, respectively. 
The axion cloud is not absolutely stable. It can emit quasi-monochromatic GWs through axion pair annihilations or level transition. Then the axion mass range $10^{-13}{\rm-}10^{-12}\,{\rm eV}$ is also ruled out by the non-observation  of this GW signal  in the LIGO/Virgo data.

The above exclusions assume that the superradiant axion modes can grow unhindered to a large enough value. On the other hand, generically  axions have a self-interaction provided by the potential, e.g. the quartic coupling $\lambda_a\sim - m_a^2 N_{\rm DW}^2/f_a^2$ for Eq.~(\ref{axion_potential}),
which may interrupt the growth of the axion modes at certain point. 
If this self-interaction is strong enough, the axion cloud can collapse into the BH before a significant portion of the angular momentum is extracted, creating a GW/axion burst called the bosenova \cite{1004.3558}. Numerical simulations and perturbative estimations indicate that the bosenova occurs 
for
$f_a/N_{\rm DW} \lesssim   \left({\Delta J}/{G_NM_{\rm BH}^2}\right)^{1/2} \alpha_g^{3/2} M_P$, 
where $\Delta J$ ($\leq G_N M_{\rm BH}^2$) is the extracted angular momentum of the BH before the collapse. Then,
parts of the axion cloud are blown away in the form of GW/axion burst, and the axion cloud will grow again and collapse at some point. If this cycle is repeated many times during the dynamical time scale of the BH, the whole process may take away a large portion of the BH angular momentum. 
Also, even before the bosenova is triggered, the axion quartic coupling may prevent the exponential growth of the axion cloud and cause early saturation by an efficient energy transfer from the superradiant mode to the damped mode \cite{1604.06422,2011.11646}. 
Considering those roles of self-interaction, the condition for extracting a sizable amount of the BH angular momentum is approximately given by \cite{1411.2263,2009.07206,2011.11646}
\dis{
  {\cal O}(10^{15}{\rm-}10^{16}\,{\rm GeV})\min\left[1, \left({10^{-18}\, \textrm{eV}}/{m_a} \right)^{1/2} \right]  \lesssim f_a/N_{\rm DW}.
} 
This effect is taken into account in the pink-shaded region in Fig.~\ref{gravity}, which is  excluded by the BH superradiance.
Even when the axion quartic coupling is weak enough to satisfy the above bound, yet the BH superradiance bounds can be avoided if the axion has a hierarchically large coupling to other fields causing an early saturation of the growth of the axion cloud \cite{1910.06308,2004.12326}.  This can be  due to an axion-dark photon coupling $g_{a\gamma^\prime}\gg g_{a\Lambda}=N_{\rm DW}/f_a$ which is technically natural and
  can arise from the clockwork mechanism discussed in Sec.~\ref{sec:landscape}.

For ultralight DE-like or DM axions, 
the gravitational probes can reach to a nearly-Planckian value of
$f_a/N_{\rm DW}$. The corresponding coupling $g_{a\Lambda}=N_{\rm DW}/f_a$ might be significantly weaker 
than the axion-instanton coupling $g_{a{\rm WGC}}$ which is bounded as Eq.~(\ref{wgc_bound2}) by the WGC.   
To see this, we use Eq.~(\ref{wgc_bound}) and Eq.~(\ref{wgc_gravitino}) with $m_{3/2}\gtrsim 10$ TeV for the bound on $g_{a\rm WGC}$, Eq.~(\ref{alp_dm}) with $\Theta_{\rm in}=1$ for the ALP DM density determined by $m_a$ and $g_{a\Lambda}=N_{\rm DW}/f_a$, and
display the resulting $g_{a{\rm WGC}}/g_{a\Lambda}$ in  Fig.~\ref{gravity}. 
The observation of a signal below the line of $g_{a{\rm WGC}}/g_{a\Lambda}\geq 1$ would challenge the assumption that the axion potential is dominantly generated by the WGC instanton.  However it is yet compatible with the axion WGC which just implies the existence of certain instanton with a coupling $g_{a\rm WGC}\gtrsim S_I/M_P$ while leaving  $g_{a\Lambda}< g_{a\rm WGC}$ as an open possibility.
 Our result shows that future CMB and PTA observations can probe the region with $g_{a{\rm WGC}}/g_{a\Lambda}\geq 10$.

\section{Axions with hierarchical couplings} \label{sec:hierarchy}

An interesting feature of axions is that there can be technically natural hierarchies
among the couplings of a given axion. Moreover, those hierarchies  can have
observable consequences. 
 We have seen that, since
axions are periodic field, their PQ-breaking couplings are given in terms of integers; for example $g_{a\Lambda}=N_{\rm DW}/f_a$. Then the ratios of those quantized couplings  do not receive quantum correction.
As a consequence, any hierarchy among the PQ-breaking couplings of a given axion can be technically natural, although it may require an explanation for its origin.
 Also the approximate $U(1)_{\rm PQ}$ symmetry assures that PQ-conserving couplings can be much stronger than PQ-breaking couplings without causing a fine tuning problem.
In this section, we present some examples of the well-motivated axion coupling hierarchies 
and  discuss the model building attempts to achieve those hierarchies in low energy effective theory.


\subsection{Examples}

\subsubsection{Coupling hierarchies for the laboratory search for axions} 

Our first example is the coupling hierarchies relevant for the laboratory search for axions.
We already noticed in
  Sec.~\ref{sec:ng_probe} (see Fig.~\ref{cp-conv-a} and Fig.~\ref{cp-conv-b}) that  the parameter region of ALP dark matter or the QCD axion
which is more easily accessible by the on-going or planned experiments  
has a bigger hierarchy between $g_{aX}$ ($X=\gamma, N$) and $g_{a\Lambda}$,
where $g_{a\Lambda}={N_{\rm DW}}/{f_a}$ is the coupling to generate the leading axion potential.
To see this, we use Eq.~(\ref{alp_dm}), Eq.~(\ref{1pi_coupling_list}) and Eq.~(\ref{qcd_axion_mass}), 
 and find
\dis{
\label{ma_gX}
\hskip -0.3cm g_{aX}\,
\sim
\left\{\begin{array}{ll} 3\times 10^{-12} R_X \left(\frac{m_a}{\rm eV}\right)^{1/4} \left(\frac{0.1 \Theta_{\rm in}^2}{\Omega_a h^2}\right)^{1/2}\, {\rm GeV}^{-1}\quad \mbox{(ALP dark matter)}  \\ 10^{-13} R_X\left(\frac{m_{a}}{\mu \rm eV}\right)\, {\rm GeV}^{-1}\quad
  \mbox{(QCD axion)}\end{array},
\right. 
}
where
\bea
R_{\gamma} =\frac{\alpha_{\rm em}}{2\pi}\frac{{\rm max}(c_G,c_W, c_B)}{N_{\rm DW}}, \quad R_N=m_N\frac{{\rm max}(c_G, c_q)}{N_{\rm DW}}.
\eea
The above result shows that for a given axion mass the corresponding $g_{aX}$ 
has a bigger value for bigger $R_X$, so is easier to be detected.
These correlations between  $g_{aX}$ and $m_a$ are displayed by the pink (ALP dark matter) and blue (QCD axion) lines
in Fig.~\ref{cp-conv-a} and Fig.~\ref{cp-conv-b}  for three different values of the hierarchy factor $R_X$ determined by $c_\gamma/N_{\rm DW}=(c_W+c_B)/N_{\rm DW}=1,10^2,10^4$ and $c_q/N_{\rm DW}=1,10^2,10^4$. For ALP dark matter, $\rho_a=\rho_{\rm DM}$ and $\Theta_{\rm in}=1$ are also assumed.

Another interesting possibility is that one of $g_{aX}$ ($X=\gamma, N, e$) is much stronger than the others,  e.g.  
i) the photophilic limit with $c_{B,W}\gg c_{G,q}$ \cite{1512.05295, 1611.09855, 1709.06085, 2008.02279, 2010.15846},
ii)
the nucleophilic limit with $c_q\gg c_{G,W,B}$ \cite{1803.07575, 2008.02279, 2010.15846},
and iii) the leptophilic limit with $c_e\gg c_{G,q}$ \cite{2007.08834, 2010.15846}.
Since most of the axion search experiments are designed to be sensitive to one specific  coupling,  these limits also might be more easily probed by the on-going or planned experiments.
Yet, the natural range of these parameter hierarchies
 are  limited
by the renormalization group mixings between
$c_X$ ($X=G,W,B$) and $c_\Psi$ ($\Psi=q,e$) induced by the SM interactions,  for instance Eq.~(\ref{rg_psi}).
Also, at scales around the nucleon or electron mass, the 1PI axion-photon coupling $g_{a\gamma}$ 
receives a threshold correction $\delta g_{a\gamma}={\cal O}\Big(\frac{\alpha_{\rm em}}{2\pi}\frac{ p_a^2}{m_\Psi^2}\frac{g_{a\Psi}}{m_\Psi}\Big)$
from the axion-fermion coupling $g_{a\Psi}$ ($\Psi=e, N$), where $p_a$ denotes the axion 4-momentum.
Taking those quantum corrections into account, we find the following naturalness bounds on the possible hierarchies among $g_{aX}$ ($X=\gamma, N, e$):
\bea
&& \hskip -0.6cm \mbox{Photophilic limit:}  \quad  1\, \ll \, \frac{c_{W,B}}{c_{q,G}} \,\sim\,
\frac{2\pi m_N}{\alpha_{\rm em}}\frac{g_{a\gamma}}{g_{aN}}\,\lesssim \,{\cal O}\left(\frac{4\pi^2}{\alpha_{\rm em}^2}\frac{1}{\ln (f_a/m_N)}\right),\nonumber \\
&& \hskip -0.6cm \mbox{Nucleophilic limit:} \quad 1 \, \ll\,\frac{c_q}{c_{B,W,G}+{\cal O}(p_a^2c_q/m_N^2)} \,\sim\, \frac{\alpha_{\rm em}}{2\pi m_N}\frac{g_{aN}}{g_{a\gamma}}\,\lesssim\, {\cal O}\left(\frac{m_N^2}{p_a^2}\right),\nonumber \\
&& \hskip -0.6cm  \mbox{Leptophilic limit:}\quad 1 \, \ll\,\frac{c_e}{c_{G,q}}\,\sim\, 
 \frac{m_N}{m_e}\frac{g_{ae}}{g_{aN}} \,\lesssim \,{\cal O}\left(\frac{4\pi^2}{\alpha_{\rm em}^2}\frac{1}{\ln (f_a/m_N)}\right).  
\eea



\subsubsection{Coupling hierarchy for the relaxion} \label{subsec:relaxion_coupling}

In the relaxion solution to the weak scale hierarchy problem, which was briefly described  in
Sec.~\ref{sec:relaxion}, 
the technically \emph{unnatural} hierarchy between $v=246$ GeV and the Higgs mass cutoff scale $\Lambda_H$ is traded for a technically \emph{natural} but much bigger  hierarchy between the relaxion couplings. For instance, for the original  model with $\Lambda_{\rm br}\lesssim v$ \cite{1504.07551}, the relaxion stabilization condition 
Eq.~(\ref{rel_hie_1})
 leads to
\dis{
\frac{g_{a{\rm br}}}{g_{a\Lambda}}=
\frac{N_{\rm br}}{N_{\rm DW}}\sim \frac{\Lambda^4}{\Lambda_{\rm br}^4}  \gtrsim \frac{\Lambda_H^4}{v^4} \label{rel_hie_2}
}
where $g_{a\Lambda}=N_{\rm DW}/f_a$ and $g_{a{\rm br}}=N_{\rm br}/f_a$ are the relaxion couplings to generate  $V_0$ and $V_{\rm br}$ in the relaxion potential Eq.~(\ref{relax_pot}). 


Similar hierarchy is required for different models which exploit different mechanisms 
to dissipate away the relaxion kinetic energy \cite{1507.07525, 1607.01786, 1805.04543, 1811.06520, 1904.02545, 1909.07706, 1911.08473}.
Here we consider two examples in which the dissipation  is dominated by bosonic particle production \cite{1607.01786, 1911.08473}. 
As the Hubble friction is negligible in these examples, the relaxion initial velocity can be greater than the barrier height as $\dot{a}_0 > \Lambda_{\rm br}^2$. Yet the relaxion can efficiently lose its kinetic energy by developing non-homogeneous modes (i.e. producing relaxion quanta),  which is dubbed ``relaxion fragmentation" \cite{1911.08473}. The relaxion will eventually be trapped at a local potential minimum when its velocity drops below the barrier height.   Successful implementation of the relaxion fragmentation requires
 \dis{
 \frac{g_{a{\rm br}}}{g_{a\Lambda}}=
\frac{N_{\rm br}}{N_{\rm DW}} \sim \max\left[\frac{\Lambda^4}{\Lambda_{\rm br}^4}\times \frac{\dot{a}_0^2}{\Lambda_{\rm br}^4}, \,\frac{\Lambda_1^2}{v^2}  \times\frac{\dot{a}_0^4}{\Lambda_{\rm br}^8} \right] \gtrsim \frac{\Lambda_H^4}{v^4}
}
with $\dot{a}_0 \gtrsim \Lambda_{\rm br}^2$ and $\Lambda_{\rm br} \lesssim v$,
which is comparable or stronger  than the hierarchy Eq.~(\ref{rel_hie_2}).

Tachyonic production of light gauge bosons can also serve as a friction for the relaxion \cite{1607.01786, 1911.08473}. 
Contrary to the above scenarios, here the Higgs vacuum value is initially of ${\cal O}(\Lambda_H)$ and later relaxed to the observed weak scale. The relaxion is coupled to the electroweak gauge bosons as 
Eq.~(\ref{pq_breaking}), but with $c_W=-c_B$ to avoid the coupling to the photon.
Initially the $W/Z$ bosons are heavy and their production is negligible. As the Higgs vacuum value is relaxed to the observed one, these bosons  become light enough to be produced by the rolling relaxion, causing the relaxion to lose its kinetic energy and eventually stop the excursion. 
 Regarding the axion coupling hierarchy, this model requires 
\dis{
\frac{g_{a{\rm br}}}{g_{a\Lambda}}=
\frac{N_{\rm br}}{N_{\rm DW}}  >  \max\left[\frac{\Lambda_H^4}{\Lambda_{\rm br}^4}, \,\frac{\Lambda_H^2}{v^2}\right]
}
in order for the relaxion to scan the Higgs mass with enough precision. The key difference from other scenarios is that in this case
$\Lambda_{\rm br}$ is independent of the Higgs field, so can be bigger than $v$. Then
the above coupling hierarchy can be significantly weaker than those in other scenarios. On the other hand, there can be additional coupling hierarchy in this scenario due to 
the coupling to the electroweak gauge bosons: 
$c_W/8\pi^2f_a=-c_B/8\pi^2f_a\sim {v}/{\dot{a}_0}$.



\subsubsection{Coupling hierarchies for large field excursion\label{sec:large_field}}

Our next example is the coupling hierarchy associated with the axion WGC \cite{hep-th/0601001,1412.3457,1503.00795, 1503.03886,1503.04783}.
As discussed in Sec.~\ref{sec:axion_cosmology},
some well-motivated axions can have a large cosmological
field excursion $\sim 1/g_{a\Lambda}=f_a/N_{\rm DW}$ comparable to $M_P$ or even bigger. 
Since the axion-instanton coupling $g_{a{\rm WGC}}$ suggested by the WGC is  bounded  as Eq.~(\ref{wgc_bound2}),  for such axions there can be a coupling hierarchy with
\bea
g_{a{\rm WGC}}\gg 
g_{a\Lambda}.\eea  
As concrete examples, one may consider
i) axion inflation with $f_a/N_{\rm DW}\sim \sqrt{N_e}M_P$  and $m_a\sim {\cal H}_{\rm inf}/\sqrt{N_e}$, which is presented in Sec.~\ref{sec:axion_inf},
ii) axion-like dark energy (quintesence) with $f_a/N_{\rm DW}\sim M_P$ and $m_a\sim 10^{-33}$ eV, and iii) ultralight  ALP dark matter produced by the initial misalignment $\sim \Theta_{\rm in}f_a/N_{\rm DW}$, whose relic energy density is given by Eq.~(\ref{alp_dm}). 
Applying the WGC bound Eq.~(\ref{wgc_bound2}) to these cases, we find
\dis{
\hskip -0.3cm \frac{g_{a{\rm WGC}}}{g_{a\Lambda}}
\,\gtrsim\, S_I\frac{f_a/N_{\rm DW}}{M_P}
\sim
\left\{ \begin{array}{ll}
\sqrt{N_e}S_{I}\quad \mbox{(axion inflation)} \\
\quad S_{I}  \quad \quad\mbox{(quintessence axion)} \\
\Big(\frac{S_{I}}{20}\Big)\Big(\frac{10^{-22}{\rm eV}}{m_a}\Big)^{1/4} \Big(\frac{\Omega_a h^2}{0.1 \Theta_{\rm in}^2}\Big)^{1/2} \,\,\,  \mbox{(ALP dark matter)}\end{array}.\nonumber
\right. 
}


For  numerical estimate of the above coupling hierarchy, one may use  Eq.~(\ref{wgc_bound}) and Eq.~(\ref{wgc_gravitino}) together with  $m_{3/2}\sim {\cal H}_{\rm inf}\sim 10^{13}$ GeV during the early Universe inflation and $m_{3/2}\gtrsim 10$ TeV in the present universe.\footnote{This bound on $m_{3/2}$ is chosen to avoid the cosmological moduli/gravitino problems.}
One then find that roughly $g_{a{\rm WGC}}/g_{a\Lambda}\gtrsim 10^2$
for  axion inflation and  quintessence axion, and
$g_{a{\rm WGC}}/g_{a\Lambda}\gtrsim 10$ for  ALP dark matter 
with $m_a\sim 10^{-22}$ and $\Omega_ah^2\sim 0.1$ \cite{1807.00824} that
may explain the small scale problems of the cold dark matter scenario \cite{astro-ph/0003365,1610.08297}.
It is also an interesting possibility that ultralight axions with $m_a\ll 10^{-22}$ constitute a small but non-negligible fraction of dark matter, e.g. 
$\Omega_ah^2\sim 10^{-2}$, which would leave an observable imprint in future cosmological data. Such a case also results in 
$g_{a{\rm WGC}}/g_{a\Lambda}\gg 1$. In Fig.~\ref{gravity}, the value of $g_{a{\rm WGC}}/g_{a\Lambda}$ is shown 
over the parameter region where the gravitational effects of ultralight ALP dark matter can be probed by astrophysical or cosmological observations in near future.

\subsubsection{Coupling hierarchies with other cosmological or astrophysical motivations\label{sec:cos_astro}} 

In addition to the examples presented above, the coupling hierarchy $g_{a\gamma}\gg \frac{\alpha_{\rm em}}{2\pi} g_{a\Lambda}$ or a similar hierarchy between $g_{a\Lambda}$ and the axion coupling to dark gauge bosons has been exploited with a variety of different cosmological or astrophysical motivations. Because of the space limitation, here we simply list those works without any further discussion.  They include the coupling hierarchy for the magnetogenesis \cite{Turner:1987bw,hep-ph/9209238, 1802.07269}, dissipative inflation \cite{0908.4089,1608.06223}, chromonatural inflation \cite{1202.2366,1806.09621},
{reducing the abundance of the QCD axion dark matter \cite{1708.05008, 1711.06590}, production of dark photon dark matter \cite{1810.07188, 1810.07196}, ALP-photon-dark photon oscillations to explain the tentative EDGES signal of 21-cm photons \cite{1911.00532}}, avoiding the BH superradiance bounds on axion masses \cite{2004.12326}, and 
resonant photon emission from the mergers of axion stars/oscillons \cite{2009.11337}.

\subsection{Hierarchies from axion landscape \label{sec:landscape}}

In this subsection, we discuss the model building attempts to generate hierarchical axion couplings in low energy effective theory \emph{without} having hierarchical parameters in the underlying UV theory.
 Most of the coupling hierarchies discussed in the previous subsection are those
 among the quantized PQ-breaking couplings, therefore involve a large integer-valued parameter, e.g.
$c_{B,W}\gg 1$ for a photophilic axion  and $N_{\rm br}\gg 1$ for the relaxion.  
Even when all UV parameters have the values of order unity, such a
 large integer  may appear in low energy effective theory as a consequence of  introducing a large number of fields
in the UV theory \cite{hep-th/0507205,  1511.00132,  1511.01827, 1512.05295, 1610.07962, 1611.09855, 1404.6209, 1404.6923, 1404.7496, 1504.03566, 1704.07831, 1709.01080, 1709.06085, 1711.06228, 1806.09621}. 
Yet, different models can have different efficiencies, i.e. the resulting hierarchy grows differently w.r.t. the number of introduced fields.
 Here we focus on the scheme based on the axion landscape provided by the potential of many $(N_H\gg 1)$ massive  axions. We will explain that this scheme
can generate an exponential  hierarchy 
 of ${\cal O}(e^{N_H})$ among the effective couplings of light axions after the massive axions are integrated out.

Let us start with a generic effective lagrangian of multiple axions:
\bea \label{eq:start}\hskip -0.5cm 
{\cal L} = \frac{1}{2} f^2_{ij}\partial_\mu \theta^i \partial^\mu\theta^j  
-V(\theta^i)
+ \frac{k_{X i}\theta^i}{32\pi^2} F^{X \mu\nu} {\tilde F}^X_{\mu\nu} 
+   c_{\psi i} \partial_\mu \theta^i\bar\psi\bar\sigma^\mu\psi + \cdots , 
\eea 
where $\theta^i\cong \theta^i+2\pi$
($i=1,2,\cdots,N$), $k_{X i}$ are integer-valued coefficients, and
the summation over the
repeated indices $i, X, \psi$ are understood. 
The axion potential $V$ arises as a consequence of the breakdown of the PQ symmetries $[U(1)_{\rm PQ}]^N:  \theta^i\rightarrow \theta^i+c^i$  $\left(c^i={\rm constants}\right)$. 
For $p$-form zero mode axions such as Eq.~(\ref{axion_pform}), often the corresponding PQ symmetries are  broken {\it only} by non-perturbative effects \cite{Dine:1986zy, hep-ph/9902292, hep-th/0605206, 1610.08297},  
yielding $V\propto e^{-b/g_*^2}$ for some  coupling $g_*$. 
For accidental PQ symmetries, their violations are  suppressed by certain powers of $1/M_P$ \cite{Barr:1992qq, hep-th/9202003, hep-ph/9203206}, which would result in 
$V\propto (f/M_P)^{n}$ for $f\ll M_P$.
These suggest that generically axion potentials with different origin have hierarchically different size, giving 
hierarchical axion masses.
Then, at a given energy scale, some axions can be heavy enough to be frozen
at their vacuum values,  while 
the other light axions are allowed to have a dynamical evolution over their entire field range. To describe such a situation, we will analyze here a simplified model in which the axion potential has two distinct scales, $\Lambda_A$ and $\Lambda_a$, with $\Lambda_A\gg \Lambda_a$. The physics we describe will be the same for models with a range of large and small energy scales, as long as these are well separated. We write the axion potential, then, as
\bea
\label{approx}
V\,=\, V_H+V_L\,=\, -\sum_{A=1}^{N_H} \Lambda_A^4
\cos(q^A_i\theta^i)
-\sum_{a=1}^{N_L}\Lambda_a^4 \cos(p^a_i\theta^i) \quad (\Lambda_A\gg \Lambda_a),
\eea
where
$\vec q^{\,A}=(q^{A}_1,\cdots, q^A_N)$  ($A=1,\cdots, N_H$)
and $\vec p^{\, a}=(p^a_1,\cdots, p^a_N)$ ($a=1,\cdots, N_L$) with $N_L=N-N_H$
 are linearly independent integer-valued vectors.
For simplicity, here we consider the simple cosine potentials including only the dominant term for each linearly independent axion combination.  However
our discussion does not rely on this specific potential and applies for
generic periodic axion potentials.   In this system, considered in the low energy limit, 
the heavy axion combinations  $q^{A}_i\theta^i$  are frozen at the vacuum state of $V_H$. The light degrees of freedom are given by the $N_L$ axions parametrizing the directions that are not constrained by $V_H$.
 As we will see, 
 the resulting effective theory of those light axions 
can have rich structures including various coupling hierarchies and enlarged field ranges\footnote{If all terms in the potential Eq.~(\ref{approx}) have a similar size, 
the resulting structure of the axion landscape can be very different from ours. For a  discussion of such cases, see for instance \cite{1709.01080}.}.

The effective lagrangian Eq.~(\ref{eq:start}) is defined in the field basis for which the discrete symmetry to ensure the periodic nature of axions is given by
 \bea \label{discrete}
 \mathbb{Z}^N:\,\,\, 
\theta^i \, \rightarrow \, \theta^i + 2\pi \ell^i \quad (\ell^i \in \mathbb{Z}).
\eea 
However, in the regime where the axion mass hierarchies become important,
it is more convenient to use  a different field basis minimizing the mixing between the axions with hierarchically different masses. 
To find such a field basis,  we first decompose $q^A_i$ into the {Smith normal form} \cite{smith1861}:
\bea
\label{smith}
q^A_i= \sum_B \hat U^A_{\,\,B} \lambda_B \hat q^B_i,\eea
where  $\lambda_A\in \mathbb{Z}$, $\hat U=[\hat U^A_{\,\,B}] \in GL(N_H,\mathbb{Z})$, $\hat q =[  \hat q^{A}_i  ,  \hat q^a_i]
\in GL(N,\mathbb{Z})$.
Here $\hat U$ and $\hat q$ are integer-valued and invertible matrices whose inverses are  also integer-valued, and therefore $|\det \hat U|=|\det \hat q|=1$.  Then the desired field basis is obtained by the $GL(N,\mathbb{Z})$ rotation:
\bea
\theta_H^A = \hat q^A_i\theta^i, 
\quad \theta_L^a=\hat q^a_i\theta^i,
\eea
followed by the field redefinition:
\bea
\theta_H^A \,\,\rightarrow \,\, \theta_H^A, \quad  \theta^a_L\,\,\rightarrow \,\,
\theta^a_L +
\hat q^a_i (f^{-2})^{ij}\hat q^A_j (f^2_H)_{AB}\theta_H^B,
\eea
where $ f^2_{ij} (f^{-2})^{jk}=\delta^k_i$ and  
$(f^2_H)_{AB}$ is defined in Eq.~(\ref{f_HL}).
With this, 
  $\theta^i$ are parameterized as 
\bea \label{new_basis}
\theta^i 
&=& \hat n_a^i  \Big( \theta_L^a +  \hat q^a_j (f^{-2})^{jk} \hat q_k^A (f_H^2)_{AB} \theta_H^B\Big) +   \hat n^i_B \theta_H^B
 \nonumber \\
 &=&
\hat n^i_a \theta^a_L  +  (f^{-2})^{i j} \hat q^A_j  (f_H^2)_{AB} \theta_H^B,
\eea 
where the integers  $\hat n^i_a$ are given by the $N_L\times N$ submatrix of the inverse of $\hat q$:\bea
\hat q^{-1} = \left[\hat n^i_A, ~\hat n^i_a \right]^T.\eea
Note that, since an inverse is involved here, the integers $\hat n^i_a$ can be large even when all of $q^A_i$ are of order unity. This will play an important role below.

Applying the above parameterization to the lagrangian Eq.~(\ref{eq:start}), we obtain
\bea \label{eq:start2}
\hskip -0.3cm {\cal L} &=& \frac{1}{2} (f^2_L)_{ab}\partial_\mu\theta_L^a\partial^\mu\theta_L^b  
+ \frac{1}{2} (f^2_H)_{AB}\partial_\mu\theta_H^A\partial^\mu\theta_H^B\nonumber \\
 &+&  \Lambda_A^4 \cos(  
\hat U^A_B \lambda_B \theta_H^B) +\Lambda_a^4
\cos\Big( p^a_i\hat n^i_b \theta_L^b + p^a_i(f^{-2})^{ij} \hat q^A_j  (f^2_H)_{AB} \theta_H^B \Big)
\nonumber\\
&+&   
 \frac{1}{32\pi^2} \Big( k_{X i}\hat n^i_a \theta_L^a 
+ k_{X i} (f^{-2})^{ij}\hat q^A_j (f^2_H)_{AB}
\theta_H^B\Big) F^X_{\mu\nu}\tilde F^{X\mu\nu} 
\nonumber\\
&+&
\Big( c_{\psi i}\hat n^i_a \partial_\mu \theta^a_L +c_{\psi i}(f^{-2})^{ij}\hat q^A_j (f^2_H)_{AB}
\partial_\mu\theta_H^B\Big) \bar\psi \bar\sigma^\mu\psi +\cdots\eea
with the block-diagonalized kinetic metric given by
 \bea
 \label{f_HL}
 (f_L^2)_{ab}=\hat n_a^i f^2_{ij}\hat n_b^j,\quad 
 (f^2_H)_{AB} (f_H^{-2})^{BC} =\delta^A_C \ \ {\rm for}\ \
 (f^{-2}_H)^{AB}=\hat q^A_i (f^{-2})^{ij}\hat q^B_j.\eea 
It shows that in the new field basis  the mixings between the heavy axions $\theta_H^A$ and the light axions $\theta_L^a$ are  suppressed
 by $\Lambda_a^4/\Lambda_A^4\ll 1$, so can be ignored.  On the other hand,
the discrete symmetry Eq.~(\ref{discrete})  takes a more complicate form 
as $\mathbb{Z}^N=\mathbb{Z}^{N_L}\times\mathbb{Z}^{N_H}$ with
\bea
 && \hskip -1cm\mathbb{Z}^{N_L}: \,\,\, \theta_L^a \rightarrow \theta_L^a +2\pi \ell^a, 
\nonumber \\
&& \hskip -1cm\mathbb{Z}^{N_H}: \,\,\, \theta_H^A \rightarrow \theta_H^A + 2\pi \ell^A, \,\,\,\theta_L^a \rightarrow \theta_L^a -2\pi\ell^A \hat q^a_i (f^{-2})^{ij}\hat q^B_j (f_H^2)_{B A}, 
\eea
for $\ell^a,\ell^A\in \mathbb{Z}$. This implies that while  $\theta_L^a$ are $2\pi$-periodic by themselves,  $\theta_H^A$ are $2\pi$-periodic {\it modulo} the shifts of light axions $\Delta \theta_L^a=2\pi \hat q^a_i (f^{-2})^{ij}\hat q^B_j(f_H^2)_{B A}$.
As a result, 
for a gauge field strength $F^X_{\mu\nu}$ which couples to the light axions $\theta_L^a$,
the couplings of $\theta_H^A$ to $F^X_{\mu\nu}\tilde F^{X\mu\nu}$ are
 generically non-quantized, 
while the couplings of $\theta_L^a$ remain to be quantized.

Our major concern is the possibility to generate a hierarchy among the low energy couplings of the light axions $\theta_L^a$ \emph{without} introducing hierarchical parameters in the underlying UV model.
Prior to the discussion of this issue, we make a small digression to mention the coupling hierarchy of $\theta_H^A$ noticed in \cite{1709.06085}, which relies on the hierarchical structure of the original axion scales $f^2_{ij}$, so has different characteristics than the coupling hierarchies of 
{$\theta_L^a$} that will be discussed later.
If $f^2_{ij}$ have hierarchical eigenvalues and $\hat q^A_i$ is well aligned with the large eigenvalue direction while having a sizable mixing with the small eigenvalue direction, 
there exists a parameter region with
$\langle \theta_H^A|\theta^i\rangle =(f^{-2})^{ij}\hat q^B_j(f_H^2)_{B A}\gg  1$
for some $i,A$. It can enhance the coupling of $\theta_H^A$ to gauge fields  relative to the coupling to generate the leading potential.  A simple example is the photophilic
QCD axion model discussed in \cite{1709.06085}, involving two axions with 
\bea 
f^{2}_{ij} = \left(\begin{array}{cc} f_1^2  & \epsilon f_1 f_2 \\
\epsilon f_1f_2& f_2^2 \end{array}\right), \quad   \vec q^{\,A} = (0, 1),  \quad \vec k_{\gamma} = (k_{1},k_{2}),
\eea
where $f_2> \epsilon f_2 \gg f_1$,
$q^A_i=\hat q^A_i$ is the coupling of $\theta^i$ to the gluon anomaly
generating the QCD axion potential as the last term in Eq.~(\ref{V_a_eff}), and $k_{\gamma i}$ is the coupling to the photon. Then $ a(x)=f_H\theta_H^A(x)$ ($\theta_H^A= \theta_2$, $f_H\simeq f_2$) can be identified as the QCD axion with
a decay constant $f_H$,
 whose coupling  to the photon is enhanced as   
\bea
  g_{a\gamma}
  \simeq \frac{\epsilon f_2}{f_1}\frac{\alpha_{\rm em}}{2\pi}
\frac{k_1}{f_{H}}\gg  \frac{\alpha_{\rm em}}{2\pi}g_{a\Lambda}= \frac{\alpha_{\rm em}}{2\pi}\frac{1}{f_H}.
\eea
  
Let us now come back to the main issue. To examine the low energy couplings of $\theta_L^a$, we
integrate out the heavy axions $\theta_H^A$. Ignoring the small corrections of ${\cal O}(\Lambda_a^4/\Lambda_A^4)$, the vacuum solution of heavy axions is  given by $\theta_H^A=0$ for arbitrary background of $\theta_L^a$.   Applying it to Eq.~(\ref{new_basis}),
the vacuum manifiold of $V_H$ is parameterized as \bea
\theta^i=\hat n_a^i\theta_L^a\label{flat_direction},\eea
and the effective lagrangian of the corresponding light axions is given by 
\bea \label{eff_lag}
{\cal L}_{\rm eff} &=& \frac{1}{2}(f^2_L)_{ab} \partial_\mu\theta_L^a\partial^\mu\theta_L^b  
+ \Lambda_a^4 \cos(  
p^a_i \hat n^i_b \theta_L^b)
\nonumber\\
&+&   
 \frac{k_{X i}\hat n^i_a \theta_L^a }{32\pi^2} 
  F^X \tilde F^X
+
c_{\psi i}\hat n^i_a \partial_\mu \theta^a_L  \bar\psi \bar\sigma^\mu\psi +\cdots.\eea
This effective lagrangian suggests that even when the UV lagrangian
Eq. (\ref{eq:start}) does not involve any hierarchical parameter,
 e.g. all dimensionless parameters in Eq. (\ref{eq:start}) are of order unity and also
 all eigenvalues of $f^2_{ij}$ have a similar size,  \emph{if} $|\hat n_a|\gg 1$ could be obtained from
 $\vec q^{\, A}$ with $|\vec q^{\,A}|={\cal O}(1)$,
 the light axion couplings
can have a hierarchical pattern 
determined by the relative angles between $\hat n_a$ and  $\{\vec p^{\,a}, \vec k_X, \vec c_\psi\}$.
Another consequence of $|\hat n_a|\gg 1$
 is that the light axion decay constants are enlarged as 
$(f_L)^2_{ab}=\hat n^i_a f^2_{ij} \hat n^j_b\gg f^2_{ij}$.


 In fact, $|\hat n_a|\gg 1$ is a generic feature of the axion landscape scenario with $N_H\gg N_L$ \cite{1404.6209, 1404.6923}.  To see this, we first note that block diagonalization of the axion kinetic terms by  Eq.~(\ref{new_basis}) leads to the  following identity for the field space volume of the canonically normalized axions:
\bea
\label{factorization}
\det (f_L^2) \det (f_H^2) = \det (f^2).\eea
 We also find
 \bea\label{identity}
 \det (\hat n_a\cdot \hat n_b) =\det (\hat q^A\cdot\hat q^B),\eea
 where $\hat n_a = (\hat n_a^1,\cdots, \hat n_a^N)$ and $\hat q^A=(\hat q^A_1,\cdots, \hat q^A_N)$. This relation   
can be obtained from Eq. (\ref{factorization}) by taking $f^2_{ij}=f_0^2\delta_{ij}$, but is valid
independently of $f^2_{ij}$.
We can further take
 an average of this relation over the Gaussian distribution of  
$\hat q^A_i$, and find \cite{goodman1963}
 \bea
\label{exp_enhan}
\Big\langle \det ( \hat n_a\cdot \hat n_b)\Big\rangle=
 \left\langle \det ( \hat q^A\cdot \hat q^B)\right\rangle
\sim \left\langle |\hat q|^2\right\rangle^{N_H}\quad (N_H\gg 1),\eea
where $|\hat q|^2=
 \sum_{A,i} (\hat q^{A}_i)^2/N_H$.
Unless $\hat q^A_i$ have a highly specific form like $\hat q^A_i=\delta^A_i$, 
we have $|\hat q|^2>1$. Then, Eq. (\ref{exp_enhan}) implies that in the limit $N_H\gg N_L$
 the {\it generic} value of 
 $|\hat n_a|$  is exponentially large  as
 \bea
 |\hat n_a|\sim |\hat q|^{N_H/N_L} \eea
 where $|\hat q|$ ($>1$) is comparable to the typical value of  $|\vec q^{\,A}|/{\rm gcd}(\vec q^{\,A})$. With this, the light axion decay constants are exponentially enlarged as
$f_L\sim |\hat q|^{N_H/N_L}f$, and also
the effective couplings of $\theta_L^a$ can have an exponential hierarchy 
determined by the relative angles between
$\hat n_a$ and  $\{\vec p^{\, a}, \vec k_X, \vec c_\psi\}$.
We stress that this is the consequence of the periodic nature of axions which requires that 
all components of $\hat n_a$ are integer-valued.
As they represent the degenerate vacuum solution of $V_H=-\Lambda_A^4
\cos(q^A_i\theta^i)$, 
$\hat n_a$ should be orthogonal to $N_H$ linearly-independent $\vec q^{\,A}$s.
However, when $N_H\gg N_L$, it is exponentially difficult for $\hat n_a$ to point in the right direction 
with only ${\cal O}(1)$ integer-valued components, so typically $|\hat n_a|$ are forced to have exponentially large values.

So far, we have discussed the generic feature of the axion landscape scenario which can be relevant for axion coupling hierarchies. Let us now proceed
with explicit examples.
For simplicity, we consider the case of single light axion ($N_L=1$), for which \cite{1404.6209}
\bea   
\hat n^i \equiv \hat n_a^i =\frac{1}{{\rm gcd}(\vec {\cal N})}{\cal N}^i\quad 
{\rm with} \quad {\cal N}^i = \det\left(\begin{array}{ccc} \delta^i_{\, 1} & \cdots & \delta^i_{\, N}  \\ 
 q^{1}_1& \cdots &  q^{1}_N  \\
\vdots &\ddots & \vdots \\ \,
q^{N_H}_1& \cdots &\,\,   q^{N_H}_N
\end{array}\right).
\eea 
Our first example is a two axion model ($N_H=N_L=1$) realizing the mechanism for
enlarging the monotonic field range of the light axion  described in \cite{hep-ph/0409138, 1404.7127}, called the KNP alignment.
 The relevant model parameters of our example are   given by
\bea
\label{knp_model}
f^2_{ij}= f_0^2\delta_{ij},\quad  \vec q^{\,A}= \vec q=(q_1,q_2), \quad \vec p^{\,a}=\vec p=(p_1,  p_2),
\eea
where $\vec q^{\,A}$ and $\vec p^{\, a}$  determine $V_H$ and $V_L$ 
as Eq.~(\ref{approx}).  Then the canonically normalized 
light axion component and its potential are determined as
\bea
a_L=f_0\frac{{\hat n\cdot \vec \theta}}{|\hat n|} , \quad  V_L=-\Lambda_a^4\cos\left(g_{a\Lambda}a_L\right)=-\Lambda_a^4\cos \Big( \frac{\hat n\cdot \vec p}{|\hat n| f_0}a_L\Big)  \eea
 with
 \bea
\hat n=\frac{1}{{\rm gcd}(q_1,q_2)}(q_2, -q_1).
\eea
The monotonic field range   of the light axion potential, i.e.
$\Delta a_L =1/g_{a\Lambda}= f_0|\hat n|/|\hat n\cdot \vec p|$, becomes much bigger than the original scale $f_0$
in the KNP alignment limit where  
$\vec q$ and $\vec p$ 
are aligned to be nearly parallel, giving $0<|\hat n\cdot \vec p|\ll |\hat n|$.
 If any of $\vec q$ and $\vec p$ were a real-valued vector, such an alignment  could be made with $|\vec q|={\cal O}(1)$.
However for periodic axions, both
$\vec q$ and $\vec p$ are integer-valued, so $|\hat n\cdot \vec p|=0$ or $\geq1$. Accordingly, the KNP alignment requires
$|\vec q|\geq |\hat n|\gg 1$.
 This implies that
  the KNP mechanism  can yield $\Delta a_L\gg f_0$ and 
 also the coupling hierarchies, 
 but  at the expense of introducing a large integer-valued input parameter.

The original motivation of the KNP mechanism is to get $\Delta a_L> M_P$  starting from a UV model with $f_0\ll M_P$ \cite{hep-ph/0409138}. 
It has been widely discussed that such a trans-Planckian
field range obtained by the KNP mechanism may have a conflict with the axion WGC.
In those discussions applied to the model of Eq.~(\ref{knp_model}),
one often identifies $\vec g_{aH}=\vec q/f_0$ and $\vec g_{aL}=\vec p/f_0$ as the axion-instanton couplings spanning the convex hull constrained by the WGC, and find that 
the corresponding  convex hull 
 cannot satisfy Eq.~(\ref{wgc_bound2}) while giving $\Delta a_L= f_0|\hat n|/|\hat n\cdot \vec p|
 > M_P$ \cite{1503.00795,1503.03886, 1503.04783, 1504.00659, 1506.03447, 1607.06814}. 
However, as noticed in Sec.~\ref{sec:wgc}, in our viewpoint the axion WGC just
implies  the existence of certain instanton couplings whose convex hull satisfies  Eq.~(\ref{wgc_bound2}), but it does not require that the axion potential is determined dominantly by those instanton couplings \cite{1412.3457, 1503.04783, 1504.00659, 1608.06951}. 
This allows that the model Eq.~(\ref{knp_model}) gives rise to  $\Delta a_L>M_P$
without any conflict with the axion WGC. 
 Yet, the axion WGC indicates that there exists an instanton with the coupling $\vec g_{\cal I}=\vec p_{\cal I}/f_0$ \emph{not} aligned with $\vec g_{aH}=\vec q/f_0$, which makes a negligible contribution to the axion potential. Then, $\Delta a_L>M_P$
 can be obtained through the KNP alignment of $\vec g_{aH}=\vec q/f_a$ and $\vec g_{aL}=\vec p/f_a$, while the WGC condition Eq.~(\ref{wgc_bound2}) is fulfilled by the convex hull 
 spanned by $\vec g_{aH}=\vec q/f_0$ and $\vec g_{a\cal I}=\vec p_{\cal I}/f_0$.
In such case,
the two effective couplings of the light axion $a_L$, i.e. 
$g_{a\Lambda}=|\hat n\cdot \vec p|/|\hat n|f_0 < 1/M_P$ and $g_{a{\rm WGC}}=|\hat n\cdot \vec p_{\cal I}|/|\hat n|f_0> S_I/M_P$
have hierarchically different size \cite{1511.05560, 1511.07201}.



A potential drawback of the KNP model is that it requires a large integer-valued input parameter to have $|\vec q^{\,A}|\gg 1$.
Our discussion leading to Eq.~(\ref{exp_enhan}) implies that this drawback disappears in models with many axions. Statistical analysis also suggests that the scheme becomes more efficient with a larger number of axions \cite{1404.6209,1404.6923}.
In the case with  $N$ ($\gg 1$) axions, 
the model can be defined on the linear quiver \cite{hep-th/0104005} with  $N$ sites for the angular axions
$\theta^i\cong \theta^i+2\pi$. 
One may then assume that axion couplings in the quiver involve only the nearest two sites. 
Taking this assumption, the $N_H=N-1$ linearly independent couplings to generate the heavy axion potential are given by 
\bea
q^A_i = q_A\delta^A_i-q^\prime_A\delta^A_{i-1} \quad (q_A, q^\prime_A \in \mathbb{Z}).
\eea
Among such models, a particularly interesting example  is the clockwork axion model \cite{1404.6209,1511.00132, 1511.01827}
with \bea f^2_{ij}=f_0^2\delta_{ij},\quad 
q_A=q,  \quad q^\prime_A =1 \quad \forall \,\,\,A. 
\eea
The clockwork mechanism naturally arranges that
 \bea
 \label{cw_sol}
\hat n=(1,q,q^2,\cdots, q^{N-1}).\eea
Note that
in the quiver description, $\hat n_a^i$ can be interpreted as the wavefuntion profile  
of the light axion along the linear quiver, with which many of the low energy  properties of the light axion can be understood \cite{1704.07831}. 

To examine the effective couplings of the light axion 
$a_L= f_0 \hat n\cdot \vec\theta/|\hat n|$ in the clockwork axion model, let us introduce  the following additional couplings:
\bea
\vec g_{a\Lambda}=\frac{\vec p}{f_0}, \quad  \vec g_{a{\cal I}}=\frac{\vec p_{\cal I}}{f_0}, \quad
\vec g_{a X }=\frac{g_X^2}{8\pi^2}\frac{\vec k_X}{f_0},
\eea
where 
$\vec g_{a\Lambda}$ is the coupling to generate the leading potential of the light axion, $\vec g_{a{\cal I}}$ is a coupling to generate additional but subleading potential, e.g.
the coupling to the WGC instanton satisfying the bound Eq.~(\ref{wgc_bound2}) or the coupling for the
barrier potential $V_{\rm br}$ of the relaxion, and $\vec g_{a X}$ 
is the coupling to $F^X_{\mu\nu}\tilde F^{X \mu\nu}$. For the integer-valued 
$\vec p, \vec p_{\cal I}$ and $\vec k_X$, one can again assume that they involve at most the nearest two sites in the quiver.  Here we choose the simplest option involving a specific single site as it gives essentially the same result:
\bea p_i=\delta^{N_1}_i, \quad p_{{\cal I} i}=\delta^{N_2}_i, \quad k_{X i}=\delta^{N_3}_i,\eea
where $1\leq N_\ell\leq N$ ($\ell=1,2,3$).
Then the low energy effective couplings of the light axion are given by
\bea 
&&{g_{a\Lambda}}
=\frac{\hat n\cdot \vec g_{a\Lambda}}{|\hat n|} 
\,=\,\frac{q^{N_1-1}}{f_L},\quad g_{a X}
=\frac{g^2_X}{8\pi^2}\frac{\hat n\cdot \vec g_{a X}}{|\hat n|}
\,=\,\frac{g^2_X}{8\pi^2}\frac{q^{N_3-1}}{f_L}, \nonumber \\
&&
g_{a {\cal I}}  = \frac{\hat n\cdot \vec g_{a{\cal I}}}{|\hat n|} 
\,=\, \frac{q^{N_2-1}}{f_L},\quad ({\cal I}=\mbox{WGC or br}) 
\eea
with an exponentially enlarged decay constant
\bea
f_L =|\hat n|f_0 =\sqrt{\frac{q^{2N}-1}{q^{2}-1}}f_0 \sim q^{N-1} f_0.\eea

The above results  show that with appropriately chosen $1\leq N_\ell\leq N$, the model can have
an exponentially enlarged monotonic field range $\Delta a_L=1/g_{a\Lambda}\sim q^{N-N_1}f_0$ of the light axion, which can be trans-Planckian while satisfying the WGC bound Eq.~(\ref{wgc_bound2}) with
 $g_{a{\rm WGC}}={\cal O}(1/f_0)\gg g_{a\Lambda}$.  The model also exhibits a variety of  exponential hierarchies among the low energy axion couplings.
We also note that the exponentially enlarged $\Delta a_L\propto q^{N-N_1}$ 
can easily overcome the possible suppression of $f_0 \propto 1/N^p$ ($p=3-4$) which was observed for some string theory axions in \cite{1808.01282}.
It turns out that the above clockwork mechanism can be generalized to the fields with nonzero spin to generate
various parameter hierarchies in particle physics \cite{1610.07962}. It also has 
the continuum limit $N\rightarrow \infty$ leading to an extra-dimensional realization of the clockwork mechanism \cite{1610.07962, 1704.07831, 1711.06228}. Although interesting, these generalizations are beyond the scope of this review.


\section{Summary and conclusion \label{sec:conclusion}}

Axions have rich physical consequences described by a variety of coupling or scale parameters.
Those parameters include for instance i) the coupling $g_{a\Lambda}$ that generates the leading  term in the axion potential, which defines the monotonic field range of the potential and also the possible cosmological excursion of the axion field,
ii) the couplings to the SM particles, particularly those to the photon, nucleon and electron,
and iii) the axion-instanton couplings suggested by the weak gravity conjecture.
An interesting feature of the axion parameter space
 is that  
 there can be  hierarchies among the different couplings of a given  axion, 
 which have good physics motivations and at the same time are technically natural.
For instance,  the parameter regions that are most accessible
by the on-going or planned axion search experiments, including the astrophysical and cosmological observations, often correspond to  regions where the axion couplings have large hierarchies. 
The hierarchy between $g_{a\Lambda}$ and the axion couplings to the gauge fields in the SM or hidden sector has been exploited with a variety of different cosmological or astrophysical motivations.
The relaxion idea to solve the weak scale hierarchy problem essentially trades the technically unnatural hiearchy between the weak scale and the cutoff scale for a technically natural but typically bigger hierarchy between $g_{a\Lambda}$ and other relaxion couplings.
Taking the WGC bound on certain axion-instanton couplings,  a tran-Planckian (or nearly Planckian) axion field excursion may imply a hierarchy between $g_{a\Lambda}$ and the couplings to the WGC instantons. 

 In this paper, we have reviewed the recent developments in axion physics while giving particular attention to the subjects of hierarchies between axion couplings.  We first summarized the existing
observational constraints on axion couplings, as well as the projected sensitivity limits of the planned experiments, which are displayed in Figs.~\ref{cp-conv-a} - \ref{gravity}.
For comparison,
we also show in the figures
the parameter ratios which exhibit certain axion coupling hierarchies.
It is apparent that the parameter regions of greatest experimental interest require large hierarchies between the axion couplings. In the theory of axions, such hierarchies are always technically natural. But it is important to show also that these hierarchies can follow from simple model-building mechanisms. Therefore,  after presenting the examples of well-motivated axion coupling  hierarchies, 
we discussed the model building attempts to generate hierarchical axion couplings in low energy effective theory.
We have focused on a specific scheme that is
based on a landscape of many axions, and we have shown that the required coupling hierarchies appear naturally in this setting. The presence of many axions with very different scales of their potentials is common in string theory, so this scheme might be realized in certain corner in the string landscape.  The scheme is quite efficient.  It can generate an exponential hierarchy among the low energy axion couplings with appropriately chosen ${\cal O}(1)$ integer-valued parameters in the UV theory. 
The ideas connected with axion coupling hierarchies mesh in a very attractive way with the regions of the axion parameter space that will be probed in the near future by on-going and planned experiments.

\section*{Acknowledgments}
We thank M. E. Peskin for many valuable comments and suggestions for improving the draft. We also thank D. E. Kaplan, Hyungjin Kim, L. D. Luzio, D. J. E. Marsh,  G. Perez, S. Rajendran, G. Servant, G. Shiu, and P. Soler for useful comments and feedbacks. This work is supported by IBS under the project code, IBS-R018-D1.

\bibliography{axion_ref}

\providecommand{\href}[2]{#2}\begingroup\raggedright\begin{thebibliography}{100}

\bibitem{0807.3125}
J.~E. Kim and G.~Carosi, ``{Axions and the Strong CP Problem},''
  \href{http://dx.doi.org/10.1103/RevModPhys.82.557}{{\em Rev. Mod. Phys.}
  {\bfseries 82} (2010) 557--602},
  \href{http://arxiv.org/abs/0807.3125}{{\ttfamily arXiv:0807.3125 [hep-ph]}}.
  [Erratum: Rev.Mod.Phys. 91, 049902 (2019)].

\bibitem{1510.07633}
D.~J.~E. Marsh, ``{Axion Cosmology},''
  \href{http://dx.doi.org/10.1016/j.physrep.2016.06.005}{{\em Phys. Rept.}
  {\bfseries 643} (2016) 1--79},
  \href{http://arxiv.org/abs/1510.07633}{{\ttfamily arXiv:1510.07633
  [astro-ph.CO]}}.

\bibitem{2003.01100}
L.~Di~Luzio, M.~Giannotti, E.~Nardi, and L.~Visinelli, ``{The landscape of QCD
  axion models},'' \href{http://dx.doi.org/10.1016/j.physrep.2020.06.002}{{\em
  Phys. Rept.} {\bfseries 870} (2020) 1--117},
  \href{http://arxiv.org/abs/2003.01100}{{\ttfamily arXiv:2003.01100
  [hep-ph]}}.

\bibitem{hep-th/0605206}
P.~Svrcek and E.~Witten, ``{Axions In String Theory},''
  \href{http://dx.doi.org/10.1088/1126-6708/2006/06/051}{{\em JHEP} {\bfseries
  06} (2006) 051}, \href{http://arxiv.org/abs/hep-th/0605206}{{\ttfamily
  arXiv:hep-th/0605206}}.

\bibitem{0905.4720}
A.~Arvanitaki, S.~Dimopoulos, S.~Dubovsky, N.~Kaloper, and J.~March-Russell,
  ``{String Axiverse},''
  \href{http://dx.doi.org/10.1103/PhysRevD.81.123530}{{\em Phys. Rev. D}
  {\bfseries 81} (2010) 123530},
  \href{http://arxiv.org/abs/0905.4720}{{\ttfamily arXiv:0905.4720 [hep-th]}}.

\bibitem{Peccei:1977hh}
R.~D. Peccei and H.~R. Quinn, ``{CP Conservation in the Presence of
  Instantons},'' \href{http://dx.doi.org/10.1103/PhysRevLett.38.1440}{{\em
  Phys. Rev. Lett.} {\bfseries 38} (1977) 1440--1443}.

\bibitem{Weinberg:1977ma}
S.~Weinberg, ``{A New Light Boson?},''
  \href{http://dx.doi.org/10.1103/PhysRevLett.40.223}{{\em Phys. Rev. Lett.}
  {\bfseries 40} (1978) 223--226}.

\bibitem{Wilczek:1977pj}
F.~Wilczek, ``{Problem of Strong $P$ and $T$ Invariance in the Presence of
  Instantons},'' \href{http://dx.doi.org/10.1103/PhysRevLett.40.279}{{\em Phys.
  Rev. Lett.} {\bfseries 40} (1978) 279--282}.

\bibitem{Freese:1990rb}
K.~Freese, J.~A. Frieman, and A.~V. Olinto, ``{Natural inflation with pseudo -
  Nambu-Goldstone bosons},''
  \href{http://dx.doi.org/10.1103/PhysRevLett.65.3233}{{\em Phys. Rev. Lett.}
  {\bfseries 65} (1990) 3233--3236}.

\bibitem{1504.07551}
P.~W. Graham, D.~E. Kaplan, and S.~Rajendran, ``{Cosmological Relaxation of the
  Electroweak Scale},''
  \href{http://dx.doi.org/10.1103/PhysRevLett.115.221801}{{\em Phys. Rev.
  Lett.} {\bfseries 115} no.~22, (2015) 221801},
  \href{http://arxiv.org/abs/1504.07551}{{\ttfamily arXiv:1504.07551
  [hep-ph]}}.

\bibitem{Preskill:1982cy}
J.~Preskill, M.~B. Wise, and F.~Wilczek, ``{Cosmology of the Invisible
  Axion},'' \href{http://dx.doi.org/10.1016/0370-2693(83)90637-8}{{\em Phys.
  Lett. B} {\bfseries 120} (1983) 127--132}.

\bibitem{Abbott:1982af}
L.~F. Abbott and P.~Sikivie, ``{A Cosmological Bound on the Invisible Axion},''
  \href{http://dx.doi.org/10.1016/0370-2693(83)90638-X}{{\em Phys. Lett. B}
  {\bfseries 120} (1983) 133--136}.

\bibitem{Dine:1982ah}
M.~Dine and W.~Fischler, ``{The Not So Harmless Axion},''
  \href{http://dx.doi.org/10.1016/0370-2693(83)90639-1}{{\em Phys. Lett. B}
  {\bfseries 120} (1983) 137--141}.

\bibitem{1201.5902}
P.~Arias, D.~Cadamuro, M.~Goodsell, J.~Jaeckel, J.~Redondo, and A.~Ringwald,
  ``{WISPy Cold Dark Matter},''
  \href{http://dx.doi.org/10.1088/1475-7516/2012/06/013}{{\em JCAP} {\bfseries
  06} (2012) 013}, \href{http://arxiv.org/abs/1201.5902}{{\ttfamily
  arXiv:1201.5902 [hep-ph]}}.

\bibitem{1302.1208}
M.~Meyer, D.~Horns, and M.~Raue, ``{First lower limits on the photon-axion-like
  particle coupling from very high energy gamma-ray observations},''
  \href{http://dx.doi.org/10.1103/PhysRevD.87.035027}{{\em Phys. Rev. D}
  {\bfseries 87} no.~3, (2013) 035027},
  \href{http://arxiv.org/abs/1302.1208}{{\ttfamily arXiv:1302.1208
  [astro-ph.HE]}}.

\bibitem{1605.06458}
A.~H. C\'orsico, A.~D. Romero, L.~G. Althaus, E.~Garc\'\i{}a-Berro, J.~Isern,
  S.~O. Kepler, M.~M. Miller~Bertolami, D.~J. Sullivan, and P.~Chote, ``{An
  asteroseismic constraint on the mass of the axion from the period drift of
  the pulsating DA white dwarf star L19-2},''
  \href{http://dx.doi.org/10.1088/1475-7516/2016/07/036}{{\em JCAP} {\bfseries
  07} (2016) 036}, \href{http://arxiv.org/abs/1605.06458}{{\ttfamily
  arXiv:1605.06458 [astro-ph.SR]}}.

\bibitem{1704.05189}
K.~Kohri and H.~Kodama, ``{Axion-Like Particles and Recent Observations of the
  Cosmic Infrared Background Radiation},''
  \href{http://dx.doi.org/10.1103/PhysRevD.96.051701}{{\em Phys. Rev. D}
  {\bfseries 96} no.~5, (2017) 051701},
  \href{http://arxiv.org/abs/1704.05189}{{\ttfamily arXiv:1704.05189
  [hep-ph]}}.

\bibitem{1610.00680}
K.~Choi and S.~H. Im, ``{Constraints on Relaxion Windows},''
  \href{http://dx.doi.org/10.1007/JHEP12(2016)093}{{\em JHEP} {\bfseries 12}
  (2016) 093}, \href{http://arxiv.org/abs/1610.00680}{{\ttfamily
  arXiv:1610.00680 [hep-ph]}}.

\bibitem{1610.02025}
T.~Flacke, C.~Frugiuele, E.~Fuchs, R.~S. Gupta, and G.~Perez, ``{Phenomenology
  of relaxion-Higgs mixing},''
  \href{http://dx.doi.org/10.1007/JHEP06(2017)050}{{\em JHEP} {\bfseries 06}
  (2017) 050}, \href{http://arxiv.org/abs/1610.02025}{{\ttfamily
  arXiv:1610.02025 [hep-ph]}}.

\bibitem{2004.02899}
A.~Banerjee, H.~Kim, O.~Matsedonskyi, G.~Perez, and M.~S. Safronova, ``{Probing
  the Relaxed Relaxion at the Luminosity and Precision Frontiers},''
  \href{http://dx.doi.org/10.1007/JHEP07(2020)153}{{\em JHEP} {\bfseries 07}
  (2020) 153}, \href{http://arxiv.org/abs/2004.02899}{{\ttfamily
  arXiv:2004.02899 [hep-ph]}}.

\bibitem{1602.00039}
P.~W. Graham, I.~G. Irastorza, S.~K. Lamoreaux, A.~Lindner, and K.~A. van
  Bibber, ``{Experimental Searches for the Axion and Axion-Like Particles},''
  \href{http://dx.doi.org/10.1146/annurev-nucl-102014-022120}{{\em Ann. Rev.
  Nucl. Part. Sci.} {\bfseries 65} (2015) 485--514},
  \href{http://arxiv.org/abs/1602.00039}{{\ttfamily arXiv:1602.00039
  [hep-ex]}}.

\bibitem{1801.08127}
I.~G. Irastorza and J.~Redondo, ``{New experimental approaches in the search
  for axion-like particles},''
  \href{http://dx.doi.org/10.1016/j.ppnp.2018.05.003}{{\em Prog. Part. Nucl.
  Phys.} {\bfseries 102} (2018) 89--159},
  \href{http://arxiv.org/abs/1801.08127}{{\ttfamily arXiv:1801.08127
  [hep-ph]}}.

\bibitem{2003.02206}
P.~Sikivie, ``{Invisible Axion Search Methods},''
  \href{http://dx.doi.org/10.1103/RevModPhys.93.015004}{{\em Rev. Mod. Phys.}
  {\bfseries 93} no.~1, (2021) 015004},
  \href{http://arxiv.org/abs/2003.02206}{{\ttfamily arXiv:2003.02206
  [hep-ph]}}.

\bibitem{1904.09003}
D.~Grin, M.~A. Amin, V.~Gluscevic, R.~Hlǒzek, D.~J.~E. Marsh, V.~Poulin,
  C.~Prescod-Weinstein, and T.~L. Smith, ``{Gravitational probes of ultra-light
  axions},'' \href{http://arxiv.org/abs/1904.09003}{{\ttfamily arXiv:1904.09003
  [astro-ph.CO]}}.

\bibitem{hep-th/0601001}
N.~Arkani-Hamed, L.~Motl, A.~Nicolis, and C.~Vafa, ``{The String landscape,
  black holes and gravity as the weakest force},''
  \href{http://dx.doi.org/10.1088/1126-6708/2007/06/060}{{\em JHEP} {\bfseries
  06} (2007) 060}, \href{http://arxiv.org/abs/hep-th/0601001}{{\ttfamily
  arXiv:hep-th/0601001}}.

\bibitem{1412.3457}
A.~de~la Fuente, P.~Saraswat, and R.~Sundrum, ``{Natural Inflation and Quantum
  Gravity},'' \href{http://dx.doi.org/10.1103/PhysRevLett.114.151303}{{\em
  Phys. Rev. Lett.} {\bfseries 114} no.~15, (2015) 151303},
  \href{http://arxiv.org/abs/1412.3457}{{\ttfamily arXiv:1412.3457 [hep-th]}}.

\bibitem{1503.00795}
T.~Rudelius, ``{Constraints on Axion Inflation from the Weak Gravity
  Conjecture},'' \href{http://dx.doi.org/10.1088/1475-7516/2015/9/020}{{\em
  JCAP} {\bfseries 09} (2015) 020},
  \href{http://arxiv.org/abs/1503.00795}{{\ttfamily arXiv:1503.00795
  [hep-th]}}.

\bibitem{1503.03886}
M.~Montero, A.~M. Uranga, and I.~Valenzuela, ``{Transplanckian axions!?},''
  \href{http://dx.doi.org/10.1007/JHEP08(2015)032}{{\em JHEP} {\bfseries 08}
  (2015) 032}, \href{http://arxiv.org/abs/1503.03886}{{\ttfamily
  arXiv:1503.03886 [hep-th]}}.

\bibitem{1503.04783}
J.~Brown, W.~Cottrell, G.~Shiu, and P.~Soler, ``{Fencing in the Swampland:
  Quantum Gravity Constraints on Large Field Inflation},''
  \href{http://dx.doi.org/10.1007/JHEP10(2015)023}{{\em JHEP} {\bfseries 10}
  (2015) 023}, \href{http://arxiv.org/abs/1503.04783}{{\ttfamily
  arXiv:1503.04783 [hep-th]}}.

\bibitem{1608.06951}
P.~Saraswat, ``{Weak gravity conjecture and effective field theory},''
  \href{http://dx.doi.org/10.1103/PhysRevD.95.025013}{{\em Phys. Rev. D}
  {\bfseries 95} no.~2, (2017) 025013},
  \href{http://arxiv.org/abs/1608.06951}{{\ttfamily arXiv:1608.06951
  [hep-th]}}.

\bibitem{hep-ph/0409138}
J.~E. Kim, H.~P. Nilles, and M.~Peloso, ``{Completing natural inflation},''
  \href{http://dx.doi.org/10.1088/1475-7516/2005/01/005}{{\em JCAP} {\bfseries
  01} (2005) 005}, \href{http://arxiv.org/abs/hep-ph/0409138}{{\ttfamily
  arXiv:hep-ph/0409138}}.

\bibitem{hep-th/0507205}
S.~Dimopoulos, S.~Kachru, J.~McGreevy, and J.~G. Wacker, ``{N-flation},''
  \href{http://dx.doi.org/10.1088/1475-7516/2008/08/003}{{\em JCAP} {\bfseries
  08} (2008) 003}, \href{http://arxiv.org/abs/hep-th/0507205}{{\ttfamily
  arXiv:hep-th/0507205}}.

\bibitem{1404.6209}
K.~Choi, H.~Kim, and S.~Yun, ``{Natural inflation with multiple sub-Planckian
  axions},'' \href{http://dx.doi.org/10.1103/PhysRevD.90.023545}{{\em Phys.
  Rev. D} {\bfseries 90} (2014) 023545},
  \href{http://arxiv.org/abs/1404.6209}{{\ttfamily arXiv:1404.6209 [hep-th]}}.

\bibitem{1511.00132}
K.~Choi and S.~H. Im, ``{Realizing the relaxion from multiple axions and its UV
  completion with high scale supersymmetry},''
  \href{http://dx.doi.org/10.1007/JHEP01(2016)149}{{\em JHEP} {\bfseries 01}
  (2016) 149}, \href{http://arxiv.org/abs/1511.00132}{{\ttfamily
  arXiv:1511.00132 [hep-ph]}}.

\bibitem{1511.01827}
D.~E. Kaplan and R.~Rattazzi, ``{Large field excursions and approximate
  discrete symmetries from a clockwork axion},''
  \href{http://dx.doi.org/10.1103/PhysRevD.93.085007}{{\em Phys. Rev. D}
  {\bfseries 93} no.~8, (2016) 085007},
  \href{http://arxiv.org/abs/1511.01827}{{\ttfamily arXiv:1511.01827
  [hep-ph]}}.

\bibitem{1610.07962}
G.~F. Giudice and M.~McCullough, ``{A Clockwork Theory},''
  \href{http://dx.doi.org/10.1007/JHEP02(2017)036}{{\em JHEP} {\bfseries 02}
  (2017) 036}, \href{http://arxiv.org/abs/1610.07962}{{\ttfamily
  arXiv:1610.07962 [hep-ph]}}.

\bibitem{1611.09855}
M.~Farina, D.~Pappadopulo, F.~Rompineve, and A.~Tesi, ``{The photo-philic QCD
  axion},'' \href{http://dx.doi.org/10.1007/JHEP01(2017)095}{{\em JHEP}
  {\bfseries 01} (2017) 095}, \href{http://arxiv.org/abs/1611.09855}{{\ttfamily
  arXiv:1611.09855 [hep-ph]}}.

\bibitem{1709.06085}
P.~Agrawal, J.~Fan, M.~Reece, and L.-T. Wang, ``{Experimental Targets for
  Photon Couplings of the QCD Axion},''
  \href{http://dx.doi.org/10.1007/JHEP02(2018)006}{{\em JHEP} {\bfseries 02}
  (2018) 006}, \href{http://arxiv.org/abs/1709.06085}{{\ttfamily
  arXiv:1709.06085 [hep-ph]}}.

\bibitem{Kim:1979if}
J.~E. Kim, ``{Weak Interaction Singlet and Strong CP Invariance},''
  \href{http://dx.doi.org/10.1103/PhysRevLett.43.103}{{\em Phys. Rev. Lett.}
  {\bfseries 43} (1979) 103}.

\bibitem{Shifman:1979if}
M.~A. Shifman, A.~I. Vainshtein, and V.~I. Zakharov, ``{Can Confinement Ensure
  Natural CP Invariance of Strong Interactions?},''
  \href{http://dx.doi.org/10.1016/0550-3213(80)90209-6}{{\em Nucl. Phys. B}
  {\bfseries 166} (1980) 493--506}.

\bibitem{Dine:1981rt}
M.~Dine, W.~Fischler, and M.~Srednicki, ``{A Simple Solution to the Strong CP
  Problem with a Harmless Axion},''
  \href{http://dx.doi.org/10.1016/0370-2693(81)90590-6}{{\em Phys. Lett. B}
  {\bfseries 104} (1981) 199--202}.

\bibitem{Zhitnitsky:1980tq}
A.~R. Zhitnitsky, ``{On Possible Suppression of the Axion Hadron Interactions.
  (In Russian)},'' {\em Sov. J. Nucl. Phys.} {\bfseries 31} (1980) 260.

\bibitem{Witten:1984dg}
E.~Witten, ``{Some Properties of O(32) Superstrings},''
  \href{http://dx.doi.org/10.1016/0370-2693(84)90422-2}{{\em Phys. Lett. B}
  {\bfseries 149} (1984) 351--356}.

\bibitem{hep-ph/9902292}
K.~Choi, ``{String or M theory axion as a quintessence},''
  \href{http://dx.doi.org/10.1103/PhysRevD.62.043509}{{\em Phys. Rev. D}
  {\bfseries 62} (2000) 043509},
  \href{http://arxiv.org/abs/hep-ph/9902292}{{\ttfamily arXiv:hep-ph/9902292}}.

\bibitem{hep-th/0303252}
T.~Banks, M.~Dine, P.~J. Fox, and E.~Gorbatov, ``{On the possibility of large
  axion decay constants},''
  \href{http://dx.doi.org/10.1088/1475-7516/2003/06/001}{{\em JCAP} {\bfseries
  06} (2003) 001}, \href{http://arxiv.org/abs/hep-th/0303252}{{\ttfamily
  arXiv:hep-th/0303252}}.

\bibitem{Choi:1985je}
K.~Choi and J.~E. Kim, ``{Harmful Axions in Superstring Models},''
  \href{http://dx.doi.org/10.1016/0370-2693(85)90416-2}{{\em Phys. Lett. B}
  {\bfseries 154} (1985) 393}. [Erratum: Phys.Lett.B 156, 452 (1985)].

\bibitem{1807.06211}
{\bfseries Planck} Collaboration, Y.~Akrami {\em et~al.}, ``{Planck 2018
  results. X. Constraints on inflation},''
  \href{http://dx.doi.org/10.1051/0004-6361/201833887}{{\em Astron. Astrophys.}
  {\bfseries 641} (2020) A10},
  \href{http://arxiv.org/abs/1807.06211}{{\ttfamily arXiv:1807.06211
  [astro-ph.CO]}}.

\bibitem{1606.07494}
S.~Borsanyi {\em et~al.}, ``{Calculation of the axion mass based on
  high-temperature lattice quantum chromodynamics},''
  \href{http://dx.doi.org/10.1038/nature20115}{{\em Nature} {\bfseries 539}
  no.~7627, (2016) 69--71}, \href{http://arxiv.org/abs/1606.07494}{{\ttfamily
  arXiv:1606.07494 [hep-lat]}}.

\bibitem{1610.08297}
L.~Hui, J.~P. Ostriker, S.~Tremaine, and E.~Witten, ``{Ultralight scalars as
  cosmological dark matter},''
  \href{http://dx.doi.org/10.1103/PhysRevD.95.043541}{{\em Phys. Rev. D}
  {\bfseries 95} no.~4, (2017) 043541},
  \href{http://arxiv.org/abs/1610.08297}{{\ttfamily arXiv:1610.08297
  [astro-ph.CO]}}.

\bibitem{Georgi:1986df}
H.~Georgi, D.~B. Kaplan, and L.~Randall, ``{Manifesting the Invisible Axion at
  Low-energies},'' \href{http://dx.doi.org/10.1016/0370-2693(86)90688-X}{{\em
  Phys. Lett. B} {\bfseries 169} (1986) 73--78}.

\bibitem{0803.3085}
E.~Silverstein and A.~Westphal, ``{Monodromy in the CMB: Gravity Waves and
  String Inflation},'' \href{http://dx.doi.org/10.1103/PhysRevD.78.106003}{{\em
  Phys. Rev. D} {\bfseries 78} (2008) 106003},
  \href{http://arxiv.org/abs/0803.3085}{{\ttfamily arXiv:0803.3085 [hep-th]}}.

\bibitem{Srednicki:1985xd}
M.~Srednicki, ``{Axion Couplings to Matter. 1. CP Conserving Parts},''
  \href{http://dx.doi.org/10.1016/0550-3213(85)90054-9}{{\em Nucl. Phys. B}
  {\bfseries 260} (1985) 689--700}.

\bibitem{hep-ph/9306216}
S.~Chang and K.~Choi, ``{Hadronic axion window and the big bang
  nucleosynthesis},''
  \href{http://dx.doi.org/10.1016/0370-2693(93)90656-3}{{\em Phys. Lett. B}
  {\bfseries 316} (1993) 51--56},
  \href{http://arxiv.org/abs/hep-ph/9306216}{{\ttfamily arXiv:hep-ph/9306216}}.

\bibitem{Kaplan:1985dv}
D.~B. Kaplan, ``{Opening the Axion Window},''
  \href{http://dx.doi.org/10.1016/0550-3213(85)90319-0}{{\em Nucl. Phys. B}
  {\bfseries 260} (1985) 215--226}.

\bibitem{Leutwyler:1989tn}
H.~Leutwyler and M.~A. Shifman, ``{GOLDSTONE BOSONS GENERATE PECULIAR CONFORMAL
  ANOMALIES},'' \href{http://dx.doi.org/10.1016/0370-2693(89)91730-9}{{\em
  Phys. Lett. B} {\bfseries 221} (1989) 384--388}.

\bibitem{Chivukula:1989ze}
R.~S. Chivukula, A.~G. Cohen, H.~Georgi, and A.~V. Manohar, ``{Couplings of a
  Light Higgs Boson},''
  \href{http://dx.doi.org/10.1016/0370-2693(89)91262-8}{{\em Phys. Lett. B}
  {\bfseries 222} (1989) 258--262}.

\bibitem{1511.02867}
G.~Grilli~di Cortona, E.~Hardy, J.~Pardo~Vega, and G.~Villadoro, ``{The QCD
  axion, precisely},'' \href{http://dx.doi.org/10.1007/JHEP01(2016)034}{{\em
  JHEP} {\bfseries 01} (2016) 034},
  \href{http://arxiv.org/abs/1511.02867}{{\ttfamily arXiv:1511.02867
  [hep-ph]}}.

\bibitem{2007.03319}
S.~Borsanyi, Z.~Fodor, C.~Hoelbling, L.~Lellouch, K.~K. Szabo, C.~Torrero, and
  L.~Varnhorst, ``{Ab-initio calculation of the proton and the neutron's scalar
  couplings for new physics searches},''
  \href{http://arxiv.org/abs/2007.03319}{{\ttfamily arXiv:2007.03319
  [hep-lat]}}.

\bibitem{Moody:1984ba}
J.~E. Moody and F.~Wilczek, ``{NEW MACROSCOPIC FORCES?},''
  \href{http://dx.doi.org/10.1103/PhysRevD.30.130}{{\em Phys. Rev. D}
  {\bfseries 30} (1984) 130}.

\bibitem{2006.12508}
S.~Bertolini, L.~Di~Luzio, and F.~Nesti, ``{Axion-mediated forces, CP violation
  and left-right interactions},''
  \href{http://dx.doi.org/10.1103/PhysRevLett.126.081801}{{\em Phys. Rev.
  Lett.} {\bfseries 126} no.~8, (2021) 081801},
  \href{http://arxiv.org/abs/2006.12508}{{\ttfamily arXiv:2006.12508
  [hep-ph]}}.

\bibitem{Gunion:1988mf}
J.~F. Gunion, G.~Gamberini, and S.~F. Novaes, ``{Can the Higgs Bosons of the
  Minimal Supersymmetric Model Be Detected at a Hadron Collider via Two Photon
  Decays?},'' \href{http://dx.doi.org/10.1103/PhysRevD.38.3481}{{\em Phys. Rev.
  D} {\bfseries 38} (1988) 3481}.

\bibitem{1903.06239}
E.~Palti, ``{The Swampland: Introduction and Review},''
  \href{http://dx.doi.org/10.1002/prop.201900037}{{\em Fortsch. Phys.}
  {\bfseries 67} no.~6, (2019) 1900037},
  \href{http://arxiv.org/abs/1903.06239}{{\ttfamily arXiv:1903.06239
  [hep-th]}}.

\bibitem{1402.2287}
C.~Cheung and G.~N. Remmen, ``{Naturalness and the Weak Gravity Conjecture},''
  \href{http://dx.doi.org/10.1103/PhysRevLett.113.051601}{{\em Phys. Rev.
  Lett.} {\bfseries 113} (2014) 051601},
  \href{http://arxiv.org/abs/1402.2287}{{\ttfamily arXiv:1402.2287 [hep-ph]}}.

\bibitem{Dine:1986zy}
M.~Dine, N.~Seiberg, X.~G. Wen, and E.~Witten, ``{Nonperturbative Effects on
  the String World Sheet},''
  \href{http://dx.doi.org/10.1016/0550-3213(86)90418-9}{{\em Nucl. Phys. B}
  {\bfseries 278} (1986) 769--789}.

\bibitem{1504.00659}
J.~Brown, W.~Cottrell, G.~Shiu, and P.~Soler, ``{On Axionic Field Ranges,
  Loopholes and the Weak Gravity Conjecture},''
  \href{http://dx.doi.org/10.1007/JHEP04(2016)017}{{\em JHEP} {\bfseries 04}
  (2016) 017}, \href{http://arxiv.org/abs/1504.00659}{{\ttfamily
  arXiv:1504.00659 [hep-th]}}.

\bibitem{1506.03447}
B.~Heidenreich, M.~Reece, and T.~Rudelius, ``{Weak Gravity Strongly Constrains
  Large-Field Axion Inflation},''
  \href{http://dx.doi.org/10.1007/JHEP12(2015)108}{{\em JHEP} {\bfseries 12}
  (2015) 108}, \href{http://arxiv.org/abs/1506.03447}{{\ttfamily
  arXiv:1506.03447 [hep-th]}}.

\bibitem{1607.06814}
A.~Hebecker, P.~Mangat, S.~Theisen, and L.~T. Witkowski, ``{Can Gravitational
  Instantons Really Constrain Axion Inflation?},''
  \href{http://dx.doi.org/10.1007/JHEP02(2017)097}{{\em JHEP} {\bfseries 02}
  (2017) 097}, \href{http://arxiv.org/abs/1607.06814}{{\ttfamily
  arXiv:1607.06814 [hep-th]}}.

\bibitem{0902.3251}
R.~Blumenhagen, M.~Cvetic, S.~Kachru, and T.~Weigand, ``{D-Brane Instantons in
  Type II Orientifolds},''
  \href{http://dx.doi.org/10.1146/annurev.nucl.010909.083113}{{\em Ann. Rev.
  Nucl. Part. Sci.} {\bfseries 59} (2009) 269--296},
  \href{http://arxiv.org/abs/0902.3251}{{\ttfamily arXiv:0902.3251 [hep-th]}}.

\bibitem{Sikivie:1983ip}
P.~Sikivie, ``{Experimental Tests of the Invisible Axion},''
  \href{http://dx.doi.org/10.1103/PhysRevLett.51.1415}{{\em Phys. Rev. Lett.}
  {\bfseries 51} (1983) 1415--1417}. [Erratum: Phys.Rev.Lett. 52, 695 (1984)].

\bibitem{2010.00169}
{\bfseries ADMX} Collaboration, R.~Khatiwada {\em et~al.}, ``{Axion Dark Matter
  eXperiment: Detailed Design and Operations},''
  \href{http://arxiv.org/abs/2010.00169}{{\ttfamily arXiv:2010.00169
  [astro-ph.IM]}}.

\bibitem{1910.11591}
Y.~K. Semertzidis {\em et~al.}, ``{Axion Dark Matter Research with IBS/CAPP},''
  \href{http://arxiv.org/abs/1910.11591}{{\ttfamily arXiv:1910.11591
  [physics.ins-det]}}.

\bibitem{2003.10894}
S.~Beurthey {\em et~al.}, ``{MADMAX Status Report},''
  \href{http://arxiv.org/abs/2003.10894}{{\ttfamily arXiv:2003.10894
  [physics.ins-det]}}.

\bibitem{1602.01086}
Y.~Kahn, B.~R. Safdi, and J.~Thaler, ``{Broadband and Resonant Approaches to
  Axion Dark Matter Detection},''
  \href{http://dx.doi.org/10.1103/PhysRevLett.117.141801}{{\em Phys. Rev.
  Lett.} {\bfseries 117} no.~14, (2016) 141801},
  \href{http://arxiv.org/abs/1602.01086}{{\ttfamily arXiv:1602.01086
  [hep-ph]}}.

\bibitem{1805.11753}
I.~Obata, T.~Fujita, and Y.~Michimura, ``{Optical Ring Cavity Search for Axion
  Dark Matter},'' \href{http://dx.doi.org/10.1103/PhysRevLett.121.161301}{{\em
  Phys. Rev. Lett.} {\bfseries 121} no.~16, (2018) 161301},
  \href{http://arxiv.org/abs/1805.11753}{{\ttfamily arXiv:1805.11753
  [astro-ph.CO]}}.

\bibitem{1807.08810}
D.~J.~E. Marsh, K.-C. Fong, E.~W. Lentz, L.~Smejkal, and M.~N. Ali, ``{Proposal
  to Detect Dark Matter using Axionic Topological Antiferromagnets},''
  \href{http://dx.doi.org/10.1103/PhysRevLett.123.121601}{{\em Phys. Rev.
  Lett.} {\bfseries 123} no.~12, (2019) 121601},
  \href{http://arxiv.org/abs/1807.08810}{{\ttfamily arXiv:1807.08810
  [hep-ph]}}.

\bibitem{2007.15656}
A.~Berlin, R.~T. D'Agnolo, S.~A.~R. Ellis, and K.~Zhou, ``{Heterodyne Broadband
  Detection of Axion Dark Matter},''
  \href{http://arxiv.org/abs/2007.15656}{{\ttfamily arXiv:2007.15656
  [hep-ph]}}.

\bibitem{1711.08999}
D.~F. Jackson~Kimball {\em et~al.}, ``{Overview of the Cosmic Axion Spin
  Precession Experiment (CASPEr)},''
  \href{http://dx.doi.org/10.1007/978-3-030-43761-9_13}{{\em Springer Proc.
  Phys.} {\bfseries 245} (2020) 105--121},
  \href{http://arxiv.org/abs/1711.08999}{{\ttfamily arXiv:1711.08999
  [physics.ins-det]}}.

\bibitem{1709.07852}
P.~W. Graham, D.~E. Kaplan, J.~Mardon, S.~Rajendran, W.~A. Terrano, L.~Trahms,
  and T.~Wilkason, ``{Spin Precession Experiments for Light Axionic Dark
  Matter},'' \href{http://dx.doi.org/10.1103/PhysRevD.97.055006}{{\em Phys.
  Rev. D} {\bfseries 97} no.~5, (2018) 055006},
  \href{http://arxiv.org/abs/1709.07852}{{\ttfamily arXiv:1709.07852
  [hep-ph]}}.

\bibitem{1907.03767}
I.~M. Bloch, Y.~Hochberg, E.~Kuflik, and T.~Volansky, ``{Axion-like Relics: New
  Constraints from Old Comagnetometer Data},''
  \href{http://dx.doi.org/10.1007/JHEP01(2020)167}{{\em JHEP} {\bfseries 01}
  (2020) 167}, \href{http://arxiv.org/abs/1907.03767}{{\ttfamily
  arXiv:1907.03767 [hep-ph]}}.

\bibitem{2005.11867}
P.~W. Graham, S.~Hac\i{}\"omero\u{g}lu, D.~E. Kaplan, Z.~Omarov, S.~Rajendran,
  and Y.~K. Semertzidis, ``{Storage ring probes of dark matter and dark
  energy},'' \href{http://dx.doi.org/10.1103/PhysRevD.103.055010}{{\em Phys.
  Rev. D} {\bfseries 103} no.~5, (2021) 055010},
  \href{http://arxiv.org/abs/2005.11867}{{\ttfamily arXiv:2005.11867
  [hep-ph]}}.

\bibitem{2001.08940}
{\bfseries QUAX} Collaboration, N.~Crescini {\em et~al.}, ``{Axion search with
  a quantum-limited ferromagnetic haloscope},''
  \href{http://dx.doi.org/10.1103/PhysRevLett.124.171801}{{\em Phys. Rev.
  Lett.} {\bfseries 124} no.~17, (2020) 171801},
  \href{http://arxiv.org/abs/2001.08940}{{\ttfamily arXiv:2001.08940
  [hep-ex]}}.

\bibitem{2010.03889}
C.~A.~J. O'Hare and E.~Vitagliano, ``{Cornering the axion with CP-violating
  interactions},'' \href{http://dx.doi.org/10.1103/PhysRevD.102.115026}{{\em
  Phys. Rev. D} {\bfseries 102} (2020) 115026},
  \href{http://arxiv.org/abs/2010.03889}{{\ttfamily arXiv:2010.03889
  [hep-ph]}}.

\bibitem{1611.05852}
E.~Hardy and R.~Lasenby, ``{Stellar cooling bounds on new light particles:
  plasma mixing effects},''
  \href{http://dx.doi.org/10.1007/JHEP02(2017)033}{{\em JHEP} {\bfseries 02}
  (2017) 033}, \href{http://arxiv.org/abs/1611.05852}{{\ttfamily
  arXiv:1611.05852 [hep-ph]}}.

\bibitem{1810.06467}
G.~Benato, A.~Drobizhev, S.~Rajendran, and H.~Ramani, ``{Invisible decay modes
  in nuclear gamma cascades},''
  \href{http://dx.doi.org/10.1103/PhysRevD.99.035025}{{\em Phys. Rev. D}
  {\bfseries 99} no.~3, (2019) 035025},
  \href{http://arxiv.org/abs/1810.06467}{{\ttfamily arXiv:1810.06467
  [hep-ph]}}.

\bibitem{1512.06165}
P.~W. Graham, D.~E. Kaplan, J.~Mardon, S.~Rajendran, and W.~A. Terrano, ``{Dark
  Matter Direct Detection with Accelerometers},''
  \href{http://dx.doi.org/10.1103/PhysRevD.93.075029}{{\em Phys. Rev. D}
  {\bfseries 93} no.~7, (2016) 075029},
  \href{http://arxiv.org/abs/1512.06165}{{\ttfamily arXiv:1512.06165
  [hep-ph]}}.

\bibitem{1911.11755}
L.~Badurina {\em et~al.}, ``{AION: An Atom Interferometer Observatory and
  Network},'' \href{http://dx.doi.org/10.1088/1475-7516/2020/05/011}{{\em JCAP}
  {\bfseries 05} (2020) 011}, \href{http://arxiv.org/abs/1911.11755}{{\ttfamily
  arXiv:1911.11755 [astro-ph.CO]}}.

\bibitem{1908.00802}
{\bfseries AEDGE} Collaboration, Y.~A. El-Neaj {\em et~al.}, ``{AEDGE: Atomic
  Experiment for Dark Matter and Gravity Exploration in Space},''
  \href{http://dx.doi.org/10.1140/epjqt/s40507-020-0080-0}{{\em EPJ Quant.
  Technol.} {\bfseries 7} (2020) 6},
  \href{http://arxiv.org/abs/1908.00802}{{\ttfamily arXiv:1908.00802 [gr-qc]}}.

\bibitem{1710.09387}
J.~L. Feng, I.~Galon, F.~Kling, and S.~Trojanowski, ``{Dark Higgs bosons at the
  ForwArd Search ExpeRiment},''
  \href{http://dx.doi.org/10.1103/PhysRevD.97.055034}{{\em Phys. Rev. D}
  {\bfseries 97} no.~5, (2018) 055034},
  \href{http://arxiv.org/abs/1710.09387}{{\ttfamily arXiv:1710.09387
  [hep-ph]}}.

\bibitem{1403.1290}
A.~Arvanitaki and A.~A. Geraci, ``{Resonantly Detecting Axion-Mediated Forces
  with Nuclear Magnetic Resonance},''
  \href{http://dx.doi.org/10.1103/PhysRevLett.113.161801}{{\em Phys. Rev.
  Lett.} {\bfseries 113} no.~16, (2014) 161801},
  \href{http://arxiv.org/abs/1403.1290}{{\ttfamily arXiv:1403.1290 [hep-ph]}}.

\bibitem{1710.05413}
{\bfseries ARIADNE} Collaboration, A.~A. Geraci {\em et~al.}, ``{Progress on
  the ARIADNE axion experiment},''
  \href{http://dx.doi.org/10.1007/978-3-319-92726-8_18}{{\em Springer Proc.
  Phys.} {\bfseries 211} (2018) 151--161},
  \href{http://arxiv.org/abs/1710.05413}{{\ttfamily arXiv:1710.05413
  [astro-ph.IM]}}.

\bibitem{1410.2896}
R.~Hlozek, D.~Grin, D.~J.~E. Marsh, and P.~G. Ferreira, ``{A search for
  ultralight axions using precision cosmological data},''
  \href{http://dx.doi.org/10.1103/PhysRevD.91.103512}{{\em Phys. Rev. D}
  {\bfseries 91} no.~10, (2015) 103512},
  \href{http://arxiv.org/abs/1410.2896}{{\ttfamily arXiv:1410.2896
  [astro-ph.CO]}}.

\bibitem{1708.05681}
R.~Hlozek, D.~J.~E. Marsh, and D.~Grin, ``{Using the Full Power of the Cosmic
  Microwave Background to Probe Axion Dark Matter},''
  \href{http://dx.doi.org/10.1093/mnras/sty271}{{\em Mon. Not. Roy. Astron.
  Soc.} {\bfseries 476} no.~3, (2018) 3063--3085},
  \href{http://arxiv.org/abs/1708.05681}{{\ttfamily arXiv:1708.05681
  [astro-ph.CO]}}.

\bibitem{1806.10608}
V.~Poulin, T.~L. Smith, D.~Grin, T.~Karwal, and M.~Kamionkowski,
  ``{Cosmological implications of ultralight axionlike fields},''
  \href{http://dx.doi.org/10.1103/PhysRevD.98.083525}{{\em Phys. Rev. D}
  {\bfseries 98} no.~8, (2018) 083525},
  \href{http://arxiv.org/abs/1806.10608}{{\ttfamily arXiv:1806.10608
  [astro-ph.CO]}}.

\bibitem{2003.09655}
J.~B. Bauer, D.~J.~E. Marsh, R.~Hlo\v{z}ek, H.~Padmanabhan, and A.~Lagu\"e,
  ``{Intensity Mapping as a Probe of Axion Dark Matter},''
  \href{http://dx.doi.org/10.1093/mnras/staa3300}{{\em Mon. Not. Roy. Astron.
  Soc.} {\bfseries 500} no.~3, (2020) 3162--3177},
  \href{http://arxiv.org/abs/2003.09655}{{\ttfamily arXiv:2003.09655
  [astro-ph.CO]}}.

\bibitem{1810.03227}
N.~K. Porayko {\em et~al.}, ``{Parkes Pulsar Timing Array constraints on
  ultralight scalar-field dark matter},''
  \href{http://dx.doi.org/10.1103/PhysRevD.98.102002}{{\em Phys. Rev. D}
  {\bfseries 98} no.~10, (2018) 102002},
  \href{http://arxiv.org/abs/1810.03227}{{\ttfamily arXiv:1810.03227
  [astro-ph.CO]}}.

\bibitem{1708.00015}
T.~Kobayashi, R.~Murgia, A.~De~Simone, V.~Ir\v{s}i\v{c}, and M.~Viel,
  ``{Lyman-$\alpha$ constraints on ultralight scalar dark matter: Implications
  for the early and late universe},''
  \href{http://dx.doi.org/10.1103/PhysRevD.96.123514}{{\em Phys. Rev. D}
  {\bfseries 96} no.~12, (2017) 123514},
  \href{http://arxiv.org/abs/1708.00015}{{\ttfamily arXiv:1708.00015
  [astro-ph.CO]}}.

\bibitem{1611.00036}
{\bfseries DESI} Collaboration, A.~Aghamousa {\em et~al.}, ``{The DESI
  Experiment Part I: Science,Targeting, and Survey Design},''
  \href{http://arxiv.org/abs/1611.00036}{{\ttfamily arXiv:1611.00036
  [astro-ph.IM]}}.

\bibitem{1810.08543}
D.~J.~E. Marsh and J.~C. Niemeyer, ``{Strong Constraints on Fuzzy Dark Matter
  from Ultrafaint Dwarf Galaxy Eridanus II},''
  \href{http://dx.doi.org/10.1103/PhysRevLett.123.051103}{{\em Phys. Rev.
  Lett.} {\bfseries 123} no.~5, (2019) 051103},
  \href{http://arxiv.org/abs/1810.08543}{{\ttfamily arXiv:1810.08543
  [astro-ph.CO]}}.

\bibitem{astro-ph/0003365}
W.~Hu, R.~Barkana, and A.~Gruzinov, ``{Cold and fuzzy dark matter},''
  \href{http://dx.doi.org/10.1103/PhysRevLett.85.1158}{{\em Phys. Rev. Lett.}
  {\bfseries 85} (2000) 1158--1161},
  \href{http://arxiv.org/abs/astro-ph/0003365}{{\ttfamily
  arXiv:astro-ph/0003365}}.

\bibitem{astro-ph/9810092}
U.~Seljak and M.~Zaldarriaga, ``{Measuring dark matter power spectrum from
  cosmic microwave background},''
  \href{http://dx.doi.org/10.1103/PhysRevLett.82.2636}{{\em Phys. Rev. Lett.}
  {\bfseries 82} (1999) 2636--2639},
  \href{http://arxiv.org/abs/astro-ph/9810092}{{\ttfamily
  arXiv:astro-ph/9810092}}.

\bibitem{1004.3558}
A.~Arvanitaki and S.~Dubovsky, ``{Exploring the String Axiverse with Precision
  Black Hole Physics},''
  \href{http://dx.doi.org/10.1103/PhysRevD.83.044026}{{\em Phys. Rev. D}
  {\bfseries 83} (2011) 044026},
  \href{http://arxiv.org/abs/1004.3558}{{\ttfamily arXiv:1004.3558 [hep-th]}}.

\bibitem{1501.06570}
R.~Brito, V.~Cardoso, and P.~Pani, ``{Superradiance}: {New Frontiers in Black
  Hole Physics},'' \href{http://dx.doi.org/10.1007/978-3-319-19000-6}{{\em
  Lect. Notes Phys.} {\bfseries 906} (2015) pp.1--237},
  \href{http://arxiv.org/abs/1501.06570}{{\ttfamily arXiv:1501.06570 [gr-qc]}}.

\bibitem{1604.06422}
A.~Gruzinov, ``{Black Hole Spindown by Light Bosons},''
  \href{http://arxiv.org/abs/1604.06422}{{\ttfamily arXiv:1604.06422
  [astro-ph.HE]}}.

\bibitem{2011.11646}
M.~Baryakhtar, M.~Galanis, R.~Lasenby, and O.~Simon, ``{Black hole
  superradiance of self-interacting scalar fields},''
  \href{http://arxiv.org/abs/2011.11646}{{\ttfamily arXiv:2011.11646
  [hep-ph]}}.

\bibitem{1411.2263}
A.~Arvanitaki, M.~Baryakhtar, and X.~Huang, ``{Discovering the QCD Axion with
  Black Holes and Gravitational Waves},''
  \href{http://dx.doi.org/10.1103/PhysRevD.91.084011}{{\em Phys. Rev. D}
  {\bfseries 91} no.~8, (2015) 084011},
  \href{http://arxiv.org/abs/1411.2263}{{\ttfamily arXiv:1411.2263 [hep-ph]}}.

\bibitem{2009.07206}
M.~J. Stott, ``{Ultralight Bosonic Field Mass Bounds from Astrophysical Black
  Hole Spin},'' \href{http://arxiv.org/abs/2009.07206}{{\ttfamily
  arXiv:2009.07206 [hep-ph]}}.

\bibitem{1910.06308}
H.~Fukuda and K.~Nakayama, ``{Aspects of Nonlinear Effect on Black Hole
  Superradiance},'' \href{http://dx.doi.org/10.1007/JHEP01(2020)128}{{\em JHEP}
  {\bfseries 01} (2020) 128}, \href{http://arxiv.org/abs/1910.06308}{{\ttfamily
  arXiv:1910.06308 [hep-ph]}}.

\bibitem{2004.12326}
A.~Mathur, S.~Rajendran, and E.~H. Tanin, ``{Clockwork mechanism to remove
  superradiance limits},''
  \href{http://dx.doi.org/10.1103/PhysRevD.102.055015}{{\em Phys. Rev. D}
  {\bfseries 102} no.~5, (2020) 055015},
  \href{http://arxiv.org/abs/2004.12326}{{\ttfamily arXiv:2004.12326
  [hep-ph]}}.

\bibitem{1512.05295}
T.~Higaki, K.~S. Jeong, N.~Kitajima, and F.~Takahashi, ``{The QCD Axion from
  Aligned Axions and Diphoton Excess},''
  \href{http://dx.doi.org/10.1016/j.physletb.2016.01.055}{{\em Phys. Lett. B}
  {\bfseries 755} (2016) 13--16},
  \href{http://arxiv.org/abs/1512.05295}{{\ttfamily arXiv:1512.05295
  [hep-ph]}}.

\bibitem{2008.02279}
J.~A. Dror and J.~M. Leedom, ``{Cosmological Tension of Ultralight Axion Dark
  Matter and its Solutions},''
  \href{http://dx.doi.org/10.1103/PhysRevD.102.115030}{{\em Phys. Rev. D}
  {\bfseries 102} no.~11, (2020) 115030},
  \href{http://arxiv.org/abs/2008.02279}{{\ttfamily arXiv:2008.02279
  [hep-ph]}}.

\bibitem{2010.15846}
L.~Darm\'e, L.~Di~Luzio, M.~Giannotti, and E.~Nardi, ``{Selective enhancement
  of the QCD axion couplings},''
  \href{http://dx.doi.org/10.1103/PhysRevD.103.015034}{{\em Phys. Rev. D}
  {\bfseries 103} no.~1, (2021) 015034},
  \href{http://arxiv.org/abs/2010.15846}{{\ttfamily arXiv:2010.15846
  [hep-ph]}}.

\bibitem{1803.07575}
G.~Marques-Tavares and M.~Teo, ``{Light axions with large hadronic
  couplings},'' \href{http://dx.doi.org/10.1007/JHEP05(2018)180}{{\em JHEP}
  {\bfseries 05} (2018) 180}, \href{http://arxiv.org/abs/1803.07575}{{\ttfamily
  arXiv:1803.07575 [hep-ph]}}.

\bibitem{2007.08834}
C.~Han, M.~L. L\'opez-Ib\'a\~nez, A.~Melis, O.~Vives, and J.~M. Yang,
  ``{Anomaly-free leptophilic axionlike particle and its flavor violating
  tests},'' \href{http://dx.doi.org/10.1103/PhysRevD.103.035028}{{\em Phys.
  Rev. D} {\bfseries 103} no.~3, (2021) 035028},
  \href{http://arxiv.org/abs/2007.08834}{{\ttfamily arXiv:2007.08834
  [hep-ph]}}.

\bibitem{1507.07525}
E.~Hardy, ``{Electroweak relaxation from finite temperature},''
  \href{http://dx.doi.org/10.1007/JHEP11(2015)077}{{\em JHEP} {\bfseries 11}
  (2015) 077}, \href{http://arxiv.org/abs/1507.07525}{{\ttfamily
  arXiv:1507.07525 [hep-ph]}}.

\bibitem{1607.01786}
A.~Hook and G.~Marques-Tavares, ``{Relaxation from particle production},''
  \href{http://dx.doi.org/10.1007/JHEP12(2016)101}{{\em JHEP} {\bfseries 12}
  (2016) 101}, \href{http://arxiv.org/abs/1607.01786}{{\ttfamily
  arXiv:1607.01786 [hep-ph]}}.

\bibitem{1805.04543}
N.~Fonseca, E.~Morgante, and G.~Servant, ``{Higgs relaxation after
  inflation},'' \href{http://dx.doi.org/10.1007/JHEP10(2018)020}{{\em JHEP}
  {\bfseries 10} (2018) 020}, \href{http://arxiv.org/abs/1805.04543}{{\ttfamily
  arXiv:1805.04543 [hep-ph]}}.

\bibitem{1811.06520}
S.-J. Wang, ``{Paper-boat relaxion},''
  \href{http://dx.doi.org/10.1103/PhysRevD.99.095026}{{\em Phys. Rev. D}
  {\bfseries 99} no.~9, (2019) 095026},
  \href{http://arxiv.org/abs/1811.06520}{{\ttfamily arXiv:1811.06520
  [hep-ph]}}.

\bibitem{1904.02545}
M.~Ibe, Y.~Shoji, and M.~Suzuki, ``{Fast-Rolling Relaxion},''
  \href{http://dx.doi.org/10.1007/JHEP11(2019)140}{{\em JHEP} {\bfseries 11}
  (2019) 140}, \href{http://arxiv.org/abs/1904.02545}{{\ttfamily
  arXiv:1904.02545 [hep-ph]}}.

\bibitem{1909.07706}
K.~Kadota, U.~Min, M.~Son, and F.~Ye, ``{Cosmological Relaxation from Dark
  Fermion Production},'' \href{http://dx.doi.org/10.1007/JHEP02(2020)135}{{\em
  JHEP} {\bfseries 02} (2020) 135},
  \href{http://arxiv.org/abs/1909.07706}{{\ttfamily arXiv:1909.07706
  [hep-ph]}}.

\bibitem{1911.08473}
N.~Fonseca, E.~Morgante, R.~Sato, and G.~Servant, ``{Relaxion Fluctuations
  (Self-stopping Relaxion) and Overview of Relaxion Stopping Mechanisms},''
  \href{http://dx.doi.org/10.1007/JHEP05(2020)080}{{\em JHEP} {\bfseries 05}
  (2020) 080}, \href{http://arxiv.org/abs/1911.08473}{{\ttfamily
  arXiv:1911.08473 [hep-ph]}}. [Erratum: JHEP 01, 012 (2021)].

\bibitem{1807.00824}
A.~Hebecker, T.~Mikhail, and P.~Soler, ``{Euclidean wormholes, baby universes,
  and their impact on particle physics and cosmology},''
  \href{http://dx.doi.org/10.3389/fspas.2018.00035}{{\em Front. Astron. Space
  Sci.} {\bfseries 5} (2018) 35},
  \href{http://arxiv.org/abs/1807.00824}{{\ttfamily arXiv:1807.00824
  [hep-th]}}.

\bibitem{Turner:1987bw}
M.~S. Turner and L.~M. Widrow, ``{Inflation Produced, Large Scale Magnetic
  Fields},'' \href{http://dx.doi.org/10.1103/PhysRevD.37.2743}{{\em Phys. Rev.
  D} {\bfseries 37} (1988) 2743}.

\bibitem{hep-ph/9209238}
W.~D. Garretson, G.~B. Field, and S.~M. Carroll, ``{Primordial magnetic fields
  from pseudoGoldstone bosons},''
  \href{http://dx.doi.org/10.1103/PhysRevD.46.5346}{{\em Phys. Rev. D}
  {\bfseries 46} (1992) 5346--5351},
  \href{http://arxiv.org/abs/hep-ph/9209238}{{\ttfamily arXiv:hep-ph/9209238}}.

\bibitem{1802.07269}
K.~Choi, H.~Kim, and T.~Sekiguchi, ``{Late-Time Magnetogenesis Driven by
  Axionlike Particle Dark Matter and a Dark Photon},''
  \href{http://dx.doi.org/10.1103/PhysRevLett.121.031102}{{\em Phys. Rev.
  Lett.} {\bfseries 121} no.~3, (2018) 031102},
  \href{http://arxiv.org/abs/1802.07269}{{\ttfamily arXiv:1802.07269
  [hep-ph]}}.

\bibitem{0908.4089}
M.~M. Anber and L.~Sorbo, ``{Naturally inflating on steep potentials through
  electromagnetic dissipation},''
  \href{http://dx.doi.org/10.1103/PhysRevD.81.043534}{{\em Phys. Rev. D}
  {\bfseries 81} (2010) 043534},
  \href{http://arxiv.org/abs/0908.4089}{{\ttfamily arXiv:0908.4089 [hep-th]}}.

\bibitem{1608.06223}
A.~Notari and K.~Tywoniuk, ``{Dissipative Axial Inflation},''
  \href{http://dx.doi.org/10.1088/1475-7516/2016/12/038}{{\em JCAP} {\bfseries
  12} (2016) 038}, \href{http://arxiv.org/abs/1608.06223}{{\ttfamily
  arXiv:1608.06223 [hep-th]}}.

\bibitem{1202.2366}
P.~Adshead and M.~Wyman, ``{Chromo-Natural Inflation: Natural inflation on a
  steep potential with classical non-Abelian gauge fields},''
  \href{http://dx.doi.org/10.1103/PhysRevLett.108.261302}{{\em Phys. Rev.
  Lett.} {\bfseries 108} (2012) 261302},
  \href{http://arxiv.org/abs/1202.2366}{{\ttfamily arXiv:1202.2366 [hep-th]}}.

\bibitem{1806.09621}
P.~Agrawal, J.~Fan, and M.~Reece, ``{Clockwork Axions in Cosmology: Is
  Chromonatural Inflation Chrononatural?},''
  \href{http://dx.doi.org/10.1007/JHEP10(2018)193}{{\em JHEP} {\bfseries 10}
  (2018) 193}, \href{http://arxiv.org/abs/1806.09621}{{\ttfamily
  arXiv:1806.09621 [hep-th]}}.

\bibitem{1708.05008}
P.~Agrawal, G.~Marques-Tavares, and W.~Xue, ``{Opening up the QCD axion
  window},'' \href{http://dx.doi.org/10.1007/JHEP03(2018)049}{{\em JHEP}
  {\bfseries 03} (2018) 049}, \href{http://arxiv.org/abs/1708.05008}{{\ttfamily
  arXiv:1708.05008 [hep-ph]}}.

\bibitem{1711.06590}
N.~Kitajima, T.~Sekiguchi, and F.~Takahashi, ``{Cosmological abundance of the
  QCD axion coupled to hidden photons},''
  \href{http://dx.doi.org/10.1016/j.physletb.2018.04.024}{{\em Phys. Lett. B}
  {\bfseries 781} (2018) 684--687},
  \href{http://arxiv.org/abs/1711.06590}{{\ttfamily arXiv:1711.06590
  [hep-ph]}}.

\bibitem{1810.07188}
P.~Agrawal, N.~Kitajima, M.~Reece, T.~Sekiguchi, and F.~Takahashi, ``{Relic
  Abundance of Dark Photon Dark Matter},''
  \href{http://dx.doi.org/10.1016/j.physletb.2019.135136}{{\em Phys. Lett. B}
  {\bfseries 801} (2020) 135136},
  \href{http://arxiv.org/abs/1810.07188}{{\ttfamily arXiv:1810.07188
  [hep-ph]}}.

\bibitem{1810.07196}
R.~T. Co, A.~Pierce, Z.~Zhang, and Y.~Zhao, ``{Dark Photon Dark Matter Produced
  by Axion Oscillations},''
  \href{http://dx.doi.org/10.1103/PhysRevD.99.075002}{{\em Phys. Rev. D}
  {\bfseries 99} no.~7, (2019) 075002},
  \href{http://arxiv.org/abs/1810.07196}{{\ttfamily arXiv:1810.07196
  [hep-ph]}}.

\bibitem{1911.00532}
K.~Choi, H.~Seong, and S.~Yun, ``{Axion-photon-dark photon oscillation and its
  implication for 21 cm observation},''
  \href{http://dx.doi.org/10.1103/PhysRevD.102.075024}{{\em Phys. Rev. D}
  {\bfseries 102} no.~7, (2020) 075024},
  \href{http://arxiv.org/abs/1911.00532}{{\ttfamily arXiv:1911.00532
  [hep-ph]}}.

\bibitem{2009.11337}
M.~A. Amin and Z.-G. Mou, ``{Electromagnetic Bursts from Mergers of Oscillons
  in Axion-like Fields},'' \href{http://arxiv.org/abs/2009.11337}{{\ttfamily
  arXiv:2009.11337 [astro-ph.CO]}}.

\bibitem{1404.6923}
T.~Higaki and F.~Takahashi, ``{Natural and Multi-Natural Inflation in Axion
  Landscape},'' \href{http://dx.doi.org/10.1007/JHEP07(2014)074}{{\em JHEP}
  {\bfseries 07} (2014) 074}, \href{http://arxiv.org/abs/1404.6923}{{\ttfamily
  arXiv:1404.6923 [hep-th]}}.

\bibitem{1404.7496}
T.~C. Bachlechner, M.~Dias, J.~Frazer, and L.~McAllister, ``{Chaotic inflation
  with kinetic alignment of axion fields},''
  \href{http://dx.doi.org/10.1103/PhysRevD.91.023520}{{\em Phys. Rev. D}
  {\bfseries 91} no.~2, (2015) 023520},
  \href{http://arxiv.org/abs/1404.7496}{{\ttfamily arXiv:1404.7496 [hep-th]}}.

\bibitem{1504.03566}
D.~Junghans, ``{Large-Field Inflation with Multiple Axions and the Weak Gravity
  Conjecture},'' \href{http://dx.doi.org/10.1007/JHEP02(2016)128}{{\em JHEP}
  {\bfseries 02} (2016) 128}, \href{http://arxiv.org/abs/1504.03566}{{\ttfamily
  arXiv:1504.03566 [hep-th]}}.

\bibitem{1704.07831}
N.~Craig, I.~Garcia~Garcia, and D.~Sutherland, ``{Disassembling the Clockwork
  Mechanism},'' \href{http://dx.doi.org/10.1007/JHEP10(2017)018}{{\em JHEP}
  {\bfseries 10} (2017) 018}, \href{http://arxiv.org/abs/1704.07831}{{\ttfamily
  arXiv:1704.07831 [hep-ph]}}.

\bibitem{1709.01080}
T.~C. Bachlechner, K.~Eckerle, O.~Janssen, and M.~Kleban, ``{Systematics of
  Aligned Axions},'' \href{http://dx.doi.org/10.1007/JHEP11(2017)036}{{\em
  JHEP} {\bfseries 11} (2017) 036},
  \href{http://arxiv.org/abs/1709.01080}{{\ttfamily arXiv:1709.01080
  [hep-th]}}.

\bibitem{1711.06228}
K.~Choi, S.~H. Im, and C.~S. Shin, ``{General Continuum Clockwork},''
  \href{http://dx.doi.org/10.1007/JHEP07(2018)113}{{\em JHEP} {\bfseries 07}
  (2018) 113}, \href{http://arxiv.org/abs/1711.06228}{{\ttfamily
  arXiv:1711.06228 [hep-ph]}}.

\bibitem{Barr:1992qq}
S.~M. Barr and D.~Seckel, ``{Planck scale corrections to axion models},''
  \href{http://dx.doi.org/10.1103/PhysRevD.46.539}{{\em Phys. Rev. D}
  {\bfseries 46} (1992) 539--549}.

\bibitem{hep-th/9202003}
M.~Kamionkowski and J.~March-Russell, ``{Planck scale physics and the
  Peccei-Quinn mechanism},''
  \href{http://dx.doi.org/10.1016/0370-2693(92)90492-M}{{\em Phys. Lett. B}
  {\bfseries 282} (1992) 137--141},
  \href{http://arxiv.org/abs/hep-th/9202003}{{\ttfamily arXiv:hep-th/9202003}}.

\bibitem{hep-ph/9203206}
R.~Holman, S.~D.~H. Hsu, T.~W. Kephart, E.~W. Kolb, R.~Watkins, and L.~M.
  Widrow, ``{Solutions to the strong CP problem in a world with gravity},''
  \href{http://dx.doi.org/10.1016/0370-2693(92)90491-L}{{\em Phys. Lett. B}
  {\bfseries 282} (1992) 132--136},
  \href{http://arxiv.org/abs/hep-ph/9203206}{{\ttfamily arXiv:hep-ph/9203206}}.

\bibitem{smith1861}
H.~J.~S. Smith, ``{On systems of linear indeterminate equations and
  congruence},''
  \href{http://dx.doi.org/http://doi.org/10.1098/rstl.1861.0016}{{\em Phil.
  Trans. R. Soc. Lond.} {\bfseries 151 (1)} (1861) 293--326}.

\bibitem{goodman1963}
N.~R. Goodman, ``{The Distribution of the Determinant of a Complex Wishart
  Distributed Matrix},''
  \href{http://dx.doi.org/http://dx.doi.org/10.1214/aoms/1177704251}{{\em Ann.
  Math. Statist.} {\bfseries no.~1, (03, 1963)} (1963) 178--180}.

\bibitem{1404.7127}
R.~Kappl, S.~Krippendorf, and H.~P. Nilles, ``{Aligned Natural Inflation:
  Monodromies of two Axions},''
  \href{http://dx.doi.org/10.1016/j.physletb.2014.08.045}{{\em Phys. Lett. B}
  {\bfseries 737} (2014) 124--128},
  \href{http://arxiv.org/abs/1404.7127}{{\ttfamily arXiv:1404.7127 [hep-th]}}.

\bibitem{1511.05560}
R.~Kappl, H.~P. Nilles, and M.~W. Winkler, ``{Modulated Natural Inflation},''
  \href{http://dx.doi.org/10.1016/j.physletb.2015.12.073}{{\em Phys. Lett. B}
  {\bfseries 753} (2016) 653--659},
  \href{http://arxiv.org/abs/1511.05560}{{\ttfamily arXiv:1511.05560
  [hep-th]}}.

\bibitem{1511.07201}
K.~Choi and H.~Kim, ``{Aligned natural inflation with modulations},''
  \href{http://dx.doi.org/10.1016/j.physletb.2016.05.097}{{\em Phys. Lett. B}
  {\bfseries 759} (2016) 520--527},
  \href{http://arxiv.org/abs/1511.07201}{{\ttfamily arXiv:1511.07201
  [hep-th]}}.

\bibitem{hep-th/0104005}
N.~Arkani-Hamed, A.~G. Cohen, and H.~Georgi, ``{(De)constructing dimensions},''
  \href{http://dx.doi.org/10.1103/PhysRevLett.86.4757}{{\em Phys. Rev. Lett.}
  {\bfseries 86} (2001) 4757--4761},
  \href{http://arxiv.org/abs/hep-th/0104005}{{\ttfamily arXiv:hep-th/0104005}}.

\bibitem{1808.01282}
M.~Demirtas, C.~Long, L.~McAllister, and M.~Stillman, ``{The Kreuzer-Skarke
  Axiverse},'' \href{http://dx.doi.org/10.1007/JHEP04(2020)138}{{\em JHEP}
  {\bfseries 04} (2020) 138}, \href{http://arxiv.org/abs/1808.01282}{{\ttfamily
  arXiv:1808.01282 [hep-th]}}.

\end{thebibliography}\endgroup
\bibliographystyle{utphys}

\end{document}